\documentclass[atmosphere,article,accept,moreauthors,pdftex]{mdpi} 

\usepackage{xcolor}

\usepackage[margin=15pt,font=normalsize,labelfont={normalsize,bf},textfont=normalsize,format=plain]{caption}

\usepackage{tikz}
\usetikzlibrary{shapes,arrows,positioning,calc,decorations.pathreplacing}
\tikzstyle{decision} = [diamond, draw, fill=gray!20, aspect=2, 
    text width=4.5em, text badly centered, node distance=3cm, inner sep=0pt]
\tikzstyle{block} = [rectangle, draw, fill=gray!20, 
    text width=5em, rounded corners, minimum height=2em]
\tikzstyle{line} = [draw, -latex']
\tikzstyle{arrow} = [thick,->,>=stealth]

\usepackage{multicol}
\usepackage{multirow}
\usepackage{bm}

\newcommand{\diff}{\mathrm{d} }

\firstpage{1} 
\pubvolume{xx}
\issuenum{1}
\articlenumber{5}
\pubyear{2019}
\copyrightyear{2019}
\history{Received: 16 October 2019; Accepted to \textit{Atmosphere}: 26 October 2019; Published: 30 October 2019}
\updates{yes} % If there is an update available, un-comment this line

\pdfoutput=1

\Title{The Effect of Clouds as an Additional Opacity Source on the Inferred Metallicity of Giant Exoplanets}

\Author{Anna Julia Poser $^{1,}$*, Nadine Nettelmann $^{2}$ and  Ronald Redmer $^{1}$}%Ronald Redmer is different from submission system, please confirm. answer: the information in the submission system is wrong, first and last name have been changed

\AuthorNames{Anna Julia Poser, Nadine Nettelmann and Ronald Redmer}

\address{%
$^{1}$ \quad Institut für Physik, Universität Rostock, D-18051 Rostock, Germany; ronald.redmer@uni-rostock.de\\
$^{2}$ \quad Institut für Planetenforschung, Deutsches Zentrum für Luft- und Raumfahrt (DLR) Berlin, \mbox{D-12489 Berlin, Germany}; nadine.nettelmann@dlr.de
}
\corres{Correspondence: anna.poser@uni-rostock.de}

\abstract{Atmospheres regulate the planetary heat loss and therefore influence planetary
thermal evolution. Uncertainty in a giant planet's thermal state contributes to the uncertainty in the
inferred abundance of heavy elements it contains. Within an analytic atmosphere model, we here investigate the influence
that different cloud opacities and cloud depths can have {{on the metallicity of irradiated extrasolar
gas giants, which is inferred from interior models. In this work, the link between inferred metallicity and
assumed cloud properties is the thermal profile of atmosphere and interior. Therefore, we perform}} %All bold had  been deleted, please confirm. answer: we again added the bold parts
coupled atmosphere, interior, and evolution calculations. {The  atmosphere  model includes  clouds  in  a  much  simplified  manner;  it  includes  long-wave  absorption  but  neglects shortwave scattering. Within that model,} we show that optically
thick, high clouds have negligible influence, whereas deep-seated, optically very thick clouds can lead to 
warmer deep tropospheres and therefore higher bulk heavy element mass estimates.
For the young hot Jupiter WASP-10b, we find a possible enhancement in inferred metallicity of up to 10\% due to possible
silicate clouds at $\sim$0.3 bar. For WASP-39b, whose observationally derived metallicity is higher than predicted by cloudless models, we find an enhancement by at most~50\%.
{However, further work on cloud properties and their self-consistent coupling to the atmospheric structure is needed in order to reduce uncertainties in the choice of model parameter values, in particular of cloud opacities.}}


\begin{document}
%%%%%%%%%%%%%%%%%%%%%%%%%%%%%%%%%%%%%%%%%%

%%%%%%%%%%%%%%%%%%%%%%%%%%%%%%%%%%%%%%%%%%

%The order of the section titles is: Introduction,  Methods, Results, Discussion, Conclusions for these journals: atmosphere

\section{Introduction}

%%% why the metallicity is important %%%
Metallicity and core mass of giant planets contain information on protostellar disks and on the process of 
planet formation. Therefore, planetary metallicity, or bulk heavy element mass fraction $Z_p$, is an 
important parameter. Core accretion formation models that reproduce the metallicity of the solar system 
giant planets~\cite{Venturini2016} predict a rapid decrease of $Z_p$ with increasing planet mass $M_p$,
still allowing for up to $14\times$ solar ($Z_p\sim 20\%$) for a Saturn-mass planet but for less than
$3\times$ solar ($Z_P \sim 4.5\%$) for a 2~$M_{\rm Jup}$ planet.\pagebreak

%%% recent important observation: high metallicity of WASP-39b and why would we consider it to be high %%%
Recently, Wakeford et al.~(2018)~\cite{Wakeford2018} used transmission spectra to determine the metallicity 
in the atmosphere of the Saturn-mass planet WASP-39b. They retrieved a high value of $\sim$100--$200\times$ solar.
This is not only higher than the prediction from core accretion formation but also higher than the upper limit 
of $55\times$ solar for the atmospheric metallicity as inferred from structure models for this planet~\cite{Thorngren2019}. 
Moreover, for some massive giant planets such as the 3~$M_{\rm Jup}$ planet WASP-10b~\cite{Maciejewski2011a}, structure models predict a significant heavy element enrichment 
of $Z_p$ of 10\% or more~\cite{Thorngren2016}.

\begin{figure}[H]
\centering
\includegraphics[width=14 cm]{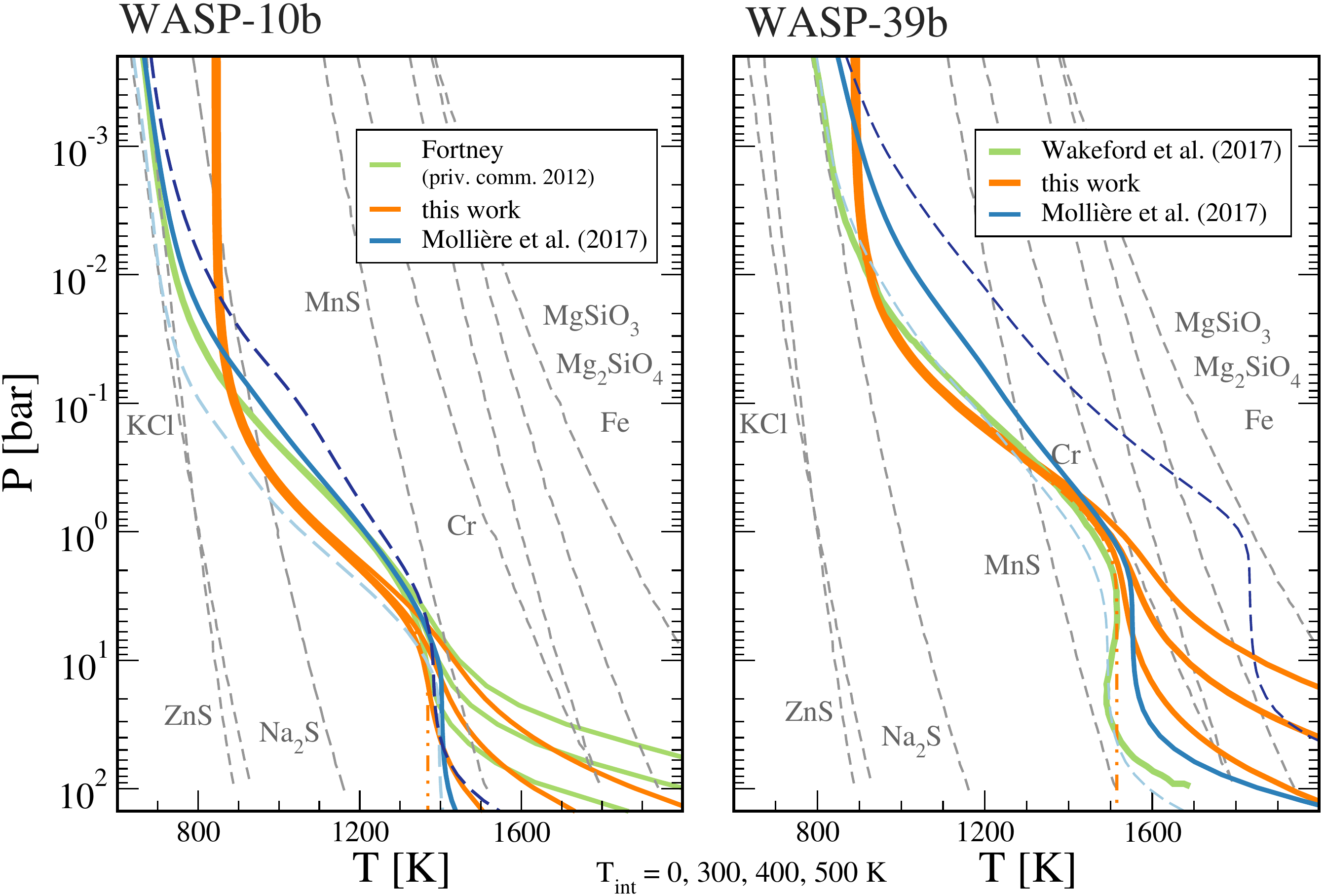}
\caption{Condensation curves (grey dashed) of some species as labeled for solar-metallicity atmospheres (taken 
from Ref.~\cite{Kataria2016}) and cloud-free $P$--$T$ profiles (solid) for  WASP-10b ({left}) and WASP-39b ({right}). 
Intersection points are possible cloud forming pressure levels. Orange $P$--$T$ profiles are our fits to the
profiles for WASP-10b from {J.~Fortney} (pers.~comm.~2012) as well as our fit to the global average profile of~\citet{Wakeford2018} for WASP-39b for different $T_{\rm int}$ values. {Additionally, we show the obtained clear profiles by~\citet{Molliere2017} for their deduced atmospheric enrichment {in} [Fe/H] (solid blue) and 10$\times$ smaller vs. larger enrichment (dashed light blue vs. dashed dark blue).} }
\label{fig:cloudfreePT} %NOTE: Please confirm only last names appear throughout. Please change [ ] to ( ) where appropriate.
\end{figure}

%%% why the atmosphere, and clouds, play an important role in this context %%%
In this paper, we pursue the possibility of uncertainty in the planet's inferred bulk metallicity due to an additional 
opacity source of limited vertical extent. We call it a cloud layer; however, we do not model any physical aspect 
of real clouds except the potential additional longwave opacity. Because of their optical properties, 
{clouds in the atmosphere are known to modify the observable transmission spectrum~\cite{Morley13gj1214} and the temperature structure of the atmosphere itself~\cite{Lines2018}.} {Clouds also influence the atmospheric scale height, which provides a direct link to the mean molecular
weight of the atmosphere~\cite{MillerRicci09}. Since the latter depends on atmospheric metallicity, its
value can be inferred from the observed transmission spectrum in combination with radiative transfer
calculations, which yield the scale heights of the observed portion of the atmosphere.  In this work, we
follow a {different approach -- inferring} %Please confirm the em dash.
the atmospheric metallicity from planetary structure models that are
primarily constrained by the observed mass, radius, and age of the star as explained below. }
In~{gaseous planets}, the radiative atmosphere transitions smoothly into the adiabatic deep interior. 
The pressure---temperature ($P$--$T$) conditions at this transition influence the internal temperatures and 
the possible intrinsic heat loss~\cite{Thorngren2019a}. Higher temperatures at a given pressure level in a fluid planet 
lead to lower densities and to expansion if not compensated for by an increase in heavy element abundance, 
an effect that is still relevant for the ice giants Uranus and Neptune~\cite{Podolak2019}. Therefore, atmospheric 
temperature profile and our inference of a planet's metallicity are strongly coupled. 
We include a cloud layer into our coupled planetary atmosphere, interior, and evolution calculations
by using the semi-analytic model of Heng et al.~(2012)~\cite{Heng2012}, which allows us to conveniently investigate the influence of assumed cloud opacity and  assumed cloud pressure level on the atmospheric $P$--$T$ profile. 
This model is applied to the two giant planets: WASP-10b and WASP-39b.
Both planets may harbor clouds since their atmospheric $P$--$T$ profiles intersect with a number of condensible
species, as shown in Figure~\ref{fig:cloudfreePT}. 

Candidates of cloud forming species for these planets 
are Na$_2$S, MnS, Cr, and silicates. {This study is not the first one to investigate the influence of cloudy and cloud-free atmospheres on the evolution of gaseous planets. Clouds have been considered in models for planets with hydrogen-dominated atmosphere before. For instance, Linder et al. (2018)~\cite{Linder2018} studied the influence on the spectra and thermal evolution of weakly irradiated exoplanets while Kurosaki et al. (2017)~\cite{Kurosaki2017} studied the influence of water clouds on the cooling of the ice giant Uranus. {For strongly irradiated hot Jupiters, Barman et al. (2001)~\cite{Barman2001} find a large heating effect in the upper atmosphere from reflection of stellar incident flux and absorption of dust grains at infrared wavelengths in comparison to clear atmospheres, with consequences on the emergent spectra, while Baraffe et al. (2003)~\cite{Baraffe2003} find a minor influence of dusty versus clear irradiated atmospheres for the luminosity evolution of hot Jupiters.}}
%%% metallicity and observations%%%
 
In Section~\ref{sec:method}, we list the relevant observed system parameters and describe our modeling approach 
for the atmosphere with a cloud layer, the interior, and the thermal evolution.
Results for WASP-10b are presented in Section~\ref{sec:resultsWASP10b} and for WASP-39b in Section~\ref{sec:resultsWASP39b}. %, respectively.
In particular, we take the $Z_p$ value of Thorngren   and   Fortney (2019)~\cite{Thorngren2019} for WASP-39b as an input parameter for our models 
and see if the high predicted atmospheric metallicity of Wakeford et al. (2018)~\cite{Wakeford2018} can be reached just
by including an additional opacity source which may mimic the effect of a cloud deck. {We compare the obtained atmospheric models with self-consistent clear and cloudy models by Molli\'{e}re et al. (2017)~\cite{Molliere2017} in Section~\ref{sec:ComparisonAtmosphere}.}
A summary is given in Section~\ref{sec:conclusions}.

%%%%%%%%%%%%%%%%%%%%%%%%%%%%%%%%%%%%%%%%%% 
%%%%%%%%%%%%%%%%%%%%%%%%%%%%%%%%%%%%%%%%%%

\section{Methods}
\label{sec:method}
\vspace{-8pt}

\subsection{Planet and Star Parameters}
\label{sec:pandsparameters}

WASP-10b is a massive ($2.96\,M_{\text{Jup}}$) and non-inflated ($T_{\rm eq}=950$ K) hot Jupiter.
Its young age of $270\pm80\,$Myr~\cite{Johnson2009,Christian2009,Maciejewski2011} makes it an interesting object 
to study planet formation and evolution. While early radius estimates predicted a rather large radius of 
$\sim$$1.27\, R_{\rm Jup}$~\cite{Christian2009}, subsequent careful analysis of the spots on the K5 dwarf host star
suggested a $20\%$ smaller planet radius of $1.02\,R_{\rm Jup}$~\cite{Maciejewski2011a}, which we use in this study. 

WASP-39b is a Saturn-mass planet ($0.28\,M_{\text{Jup}}$) with a large radius ($1.27\,R_{\text{Jup}}$) 
and therefore low density $\rho = 0.141\,\rho_{\text{Jup}}$~\cite{Faedi2011}. It is orbiting a late G-type star, 
which is smaller and, with an age of $9^{+3}_{-4}\,$Gyr, possibly older than the Sun. 
The observational parameters used here for WASP-10b and WASP-39b are listed in Table \ref{tab:starplanetparam}.

\begin{table}[H]
\caption{Stellar and planetary parameters.}
\label{tab:starplanetparam}
\centering
% \tablesize{} %% You can specify the fontsize here, e.g., \tablesize{\footnotesize}. If commented out \small will be used.
\begin{tabular}{ccc}
%\toprule
\toprule % NN
\textbf{}	& \textbf{WASP-10b}	& \textbf{WASP-39b} \boldmath{$^{5}$} \\
%\midrule
\midrule % NN
$M_P$		& $2.96^{+0.22}_{-0.17}\,M_{\text{Jup}}$ $^{1}$		& $0.28\pm0.03\,M_{\text{Jup}}$\\
$R_P$		& $1.03^{+0.077}_{-0.03}\,R_{\text{Jup}}$ $^{4}$	& $1.27\pm0.04\,R_{\text{Jup}}$\\
a		& $0.0369^{+0.0012}_{-0.0014}\,$AU	$^{1}$		& $0.0486\pm0.0005\,$AU\\
e		& $0.013\pm0.063$	 $^{3}$			& 0\\
$T_{\text{eq,A=0}}$		& $ 950^{+30}_{-26}\,$K	$^{4}$			& $ 1116^{+33}_{-32}\,$K \\%, $1030^{+30}_{-20}\,$K\\
P		& 3.09 d		& 4.05 d\\
%\midrule
\midrule % NN
$M_{\star}$		& $0.75\,M_{\text{Sun}}$ $^{2}$			& $0.93\pm0.03\,M_{\text{Sun}}$\\
$R_{\star}$		& $0.67\,R_{\text{Sun}}$ $^{4}$			& $0.895\pm0.023\,R_{\text{Sun}}$\\
$T_{\star}$		& $4675\pm100\:$K	$^{1}$				& $5400\pm150\:$K\\
age $\tau_{\star}$		& $270\pm80\,$Myr$^{3}$			& $9^{+3}_{-4}$\,Gyr\\
\bottomrule
 % NN
\end{tabular}\\
\begin{tabular}{ccc}
\multicolumn{1}{c}{\footnotesize $^{1}$ Ref.~\cite{Christian2009}, $^{2}$ Ref.~\cite{Johnson2009}, 
$^{3}$ Ref.~\cite{Maciejewski2011}, $^{4}$ Ref.~\cite{Maciejewski2011a}, $^{5}$ Ref.~\cite{Faedi2011}.}
\end{tabular}
\end{table}

Since we are interested in the effect of clouds relative to cloudless atmospheres on the inferred planet metallicity, 
we compute here planet models for a variety of cloud parameters but do not account for the observational uncertainties
in planet mass and radius. The only exception is thermal evolution calculations for WASP-39b, where we request 
its radius at present time to drop below the $1\sigma$ upper~limit.

\subsection{Interior}

To estimate the present structure of the planets, we connect the atmosphere to the interior and perform thermal 
evolution calculations. For the interior, we assume a three-layer structure of rocky core, an adiabatic, 
convective envelope, and a radiative atmosphere. Atmosphere and envelope consist of a mixture of hydrogen, 
helium and metals. Respective equations of state (EOS) are combined via the linear mixing rule. 
By heavy elements or metals, we denote all elements or molecules heavier than helium. $Z_{\text{atm}}$ and 
$Z_{\rm{env}}$ are the heavy element mass fractions in the atmosphere and envelope, respectively, which 
we assume to be equal, {$Z_{\rm{atm}}=Z_{\rm{env}}=Z$. This is an assumption, not ruling out other relations between atmospheric and envelope abundances~\cite{Molliere2017, Mordasini2016}.} The planetary bulk heavy element mass fraction
is $Z_{\rm P}=Z_{\rm{env}}M_{\text{env}}/M_{\rm P}+M_{\text{core}}/M_{\rm P}$, and $M_{\text{env}}$ and $M_{\text{core}}$ are the
masses of envelope and core. {For the solar reference metallicity we use $Z_\odot=1.5\%$~\cite{Lodders2003}}. 
For WASP-10b, we set $Z=Z_\odot$ and allow only the core mass to vary while, for WASP-39b, 
we allow also $Z$ to vary. 
%%% H, He, and EOS %%%
The helium to hydrogen mass fraction is set to the protosolar value of 
$Y=0.27$, where $Y=M_{\rm He}/(M_{\rm He}+M_{\rm H})$. For hydrogen and helium, we use the SCvH EOS 
\cite{SaumonChabrier1995}. Metals in the envelope are represented by that He-EOS scaled in density by a factor of four,
or by the ice EOS presented in~\cite{Hubbard1989a}. The rocky core obeys the pressure--density relation given in~\cite{Hubbard1989a}. 
{The density $\rho(P,T)$ is obtained from the linearly mixed EOS at the pressure $P$ and temperate $T$ 
by interpolation. We obtain the mixed EOS by adding heavy elements to the interior and the atmosphere via the linear mixing rule $\rho^{-1}(P,T)=\sum_i X_i/\rho_i(P,T)$, where $X_i$ denotes the mass fraction of component $i$ and $X_{\rm H}:=X$, $X_{\rm He}:=Y$, $X_{\rm{Z}}:=Z$~\cite{Nettelmann2011}. The density profile follows a pre-computed $P$--$T$ profile along the adiabat of the envelope.} Increasing the temperature at fixed 
pressure usually decreases the density. Lower densities in the mantle result in a larger core mass to conserve the given
planet mass. This is why the $P$--$T$ profile is so important. Otherwise, we rely on the usual structure equations for 
non-rotating, spherical giant planets as previously done in~\cite{Nettelmann2011,Fortney2010}.

%%%%%%%%%%%%%%%%%%%%%%%%%%%%%%%%%%%%%%%%%%
\subsection{Atmosphere Model with Clouds}
\label{sec:atmospheremodel}

The atmosphere model yields the atmospheric $P$--$T$ profile. We use the 1D, plane-parallel, analytical atmosphere 
model by Heng et al. (2012)~\cite{Heng2012} for hot Jupiters. 
It is based on the two-stream solution and dual band approximation, where the incoming and outgoing 
radiation fluxes are described by different frequency-averaged mean opacities. The incoming flux is represented 
by the short-wave opacity $\kappa_S$, equivalent to the opacity $\kappa_{\rm vis}$ for visual light used in
\cite{Guillot2010}, while the outgoing flux is described by the long-wave opacity $\kappa_{L}$ equivalent   
to $\kappa_{\rm th}$ in~\cite{Guillot2010} for thermal radiation. Following  \citet{Heng2012}, $\kappa_S$ is 
constant with respect to temperature and pressure while $\kappa_L$ may have a dependence on pressure. {Indeed, gas opacities significantly depend on pressure because of pressure broadening or collision induced absorption.} Cloud decks are included as an additional opacity source $\kappa_c (P)$ to the constant long-wave opacity 
$\kappa_{\text{L,0}}$ of the otherwise cloudless atmosphere,
\begin{equation}
\kappa_L (P) = \kappa_{\text{L,0}} + \kappa_c (P)~\textrm{.}	
\end{equation}

%%% T-tau %%%
{The analytic model atmosphere provides a relation between global mean temperature $T$ and longwave optical depth
$\diff\tau_L = \kappa_L \diff m$, where $m$ is column mass from top to bottom, as well as the parameter $\tau = \kappa_L m$. We call the latter here optical depth although this holds only if $\kappa_L = \rm{const.}$.}
The $T$--$\tau$ relation makes use of the Eddington coefficients $\mathcal{E}_1=1/3$ and $\mathcal{E}_2=1/2$ to close the set
of equations for the moments of radiation transfer. It reads (cf.~Equation~(31) in~\cite{Heng2012})
\begin{equation} 
	  T^4 = \frac{T_{\text{int}}^4}{4} \left(2 +  3 \int_0^m \kappa_L \diff m'  \right) 
	   + \frac{T_{\text{eq}}^4}{2} \left[ 1 + \frac{\gamma}{\sqrt{\xi}} E_2\left(\frac{\kappa_S m'}{\sqrt{\xi}}\right) + 3 \int_0^m \kappa_L  E_3\left(\frac{\kappa_S m'}{\sqrt{\xi}}\right) \diff m'   \right]~\textrm{,}	
  \label{eq:Ttau}
 \end{equation}
with $E_j(x) = \int^{\infty}_1 y^{-j} \exp(-xy)\:\diff y $ as the exponential integrals. Equation~(\ref{eq:Ttau}) depends on the cloud opacity through $\kappa_L$ and the opacity ratio $\gamma=\kappa_S/\kappa_{L}$. Furthermore, the global mean temperature $T$ depends on the intrinsic heat flux $F_{\rm int}=\sigma_{B}\,T_{\rm  int}^4$, 
which is the outgoing flux from the planet at the bottom of the atmosphere, and on the zero-albedo irradiation flux
$\sigma_{B} T_{\text{eq,\,0}}^4=\sigma_{B} T_{\star}^4(R_\star/2a)^2$ where $\sigma_{B}$ is the Stephan--Boltzmann constant.
Thus, $(1-A)^{1/4} T_{\rm eq,0}$ is the globally averaged temperature a planet of albedo $A$ would adopt if 
in radiation equilibrium with the incident flux.
Since a scattering parameter $\xi<1$ would be inconsistent with a non-uniform opacity, here $\kappa_L(P)$, we set 
$\xi=1$ (no scattering) and take scattering into account only via the albedo in $T_{\text{eq}}$, 
which we set to $A_B=0.3${~\cite{Marley1999, Gelino1999}, while noting that other work suggests smaller values (e.g.,~\cite{Sudarsky2000}). {More recently, the geometric albedo of several exoplanets has been derived from secondary eclipse data and found to be quite small, even less than 0.1~\cite{Madhusudhan2014}.} On the other hand, the Bond albedo value of Jupiter itself has recently been revised upward from its Voyager-data based value of 0.34 to the new Cassini-data based value of 0.5~\cite{Li2018}. %NOTE: Is this capitalization necessary? answer: good catch, it is not
To study the thermal evolution of irradiated giant planets as a function of uncertainty in albedo is left to future work.} For a more consistent treatment of scattering in the presence of non-uniform absorption, 
see Ref.~\cite{Heng2014}. 
For the cloud-free ($\kappa_L = \kappa_{\text{L,\,0}}$) atmosphere without scattering ($\xi$=1), Equation~(\ref{eq:Ttau}) reduces to the global average temperature profile of Guillot (2010)~\cite{Guillot2010}.
%%% P-tau %%%
{The $P$--$\tau$ relation for constant gravity $g$ and pressure-dependent longwave opacity~reads }
\begin{equation}
P= m\cdot g = (\tau/\kappa_L)\:g~\textrm{.}
\end{equation}

%%% gamma %%%
We use the cloud-free  model to constrain the parameter $\gamma$. For WASP-10b, we fit to 1D, non-gray, 
atmospheric $P$--$T$ profiles specifically calculated for this planet for different values of $T_{\rm  int}$  
\cite{Fortney2007}  (see Figure~\ref{fig:cloudfreePT}). 
For WASP-39b, we fit $\kappa_S$ and $\kappa_{\rm L,\,0}$ to the global averaged $P$--$T$ profile from Ref.~\cite{Kataria2016}
 for a $1\times$ solar composition metallicity. 
%It is based on a multi-stream radiative transfer code where non-gray opacities are 
%calculated assuming local thermodynamic and, for composition, chemical equilibirium 
%\cite{Showman2009,Kataria2015}. 
We find $\kappa_{\text{L,0}}=0.0136$ cm$^2$/g, 
$\kappa_S	= 0.002	$ cm$^2$/g ($\gamma=0.147$) for WASP-10b and  $\kappa_{\text{L,0}}=0.006$ cm$^2$/g, 
$\kappa_S	= 0.00037$ cm$^2$/g ($\gamma=0.062$) for WASP-39b. 	
{However, albeit using solar-composition models to fit our double-gray clear atmosphere, it is important to notice that an increase (or decrease) of atmospheric enrichment changes the position of the isotherm~\cite{Fortney2006,Molliere2015}. An increase in atmospheric metallicity leads to higher temperatures in the isothermal part of the atmosphere~\cite{Wakeford2017}. In the case of WASP-39b, where possibly the atmosphere is enriched by a factor of 100--200$\times$ solar value, the isotherm would be pushed to even hotter temperatures. }

\begin{figure}[H]
\centering
\includegraphics[width=10 cm]{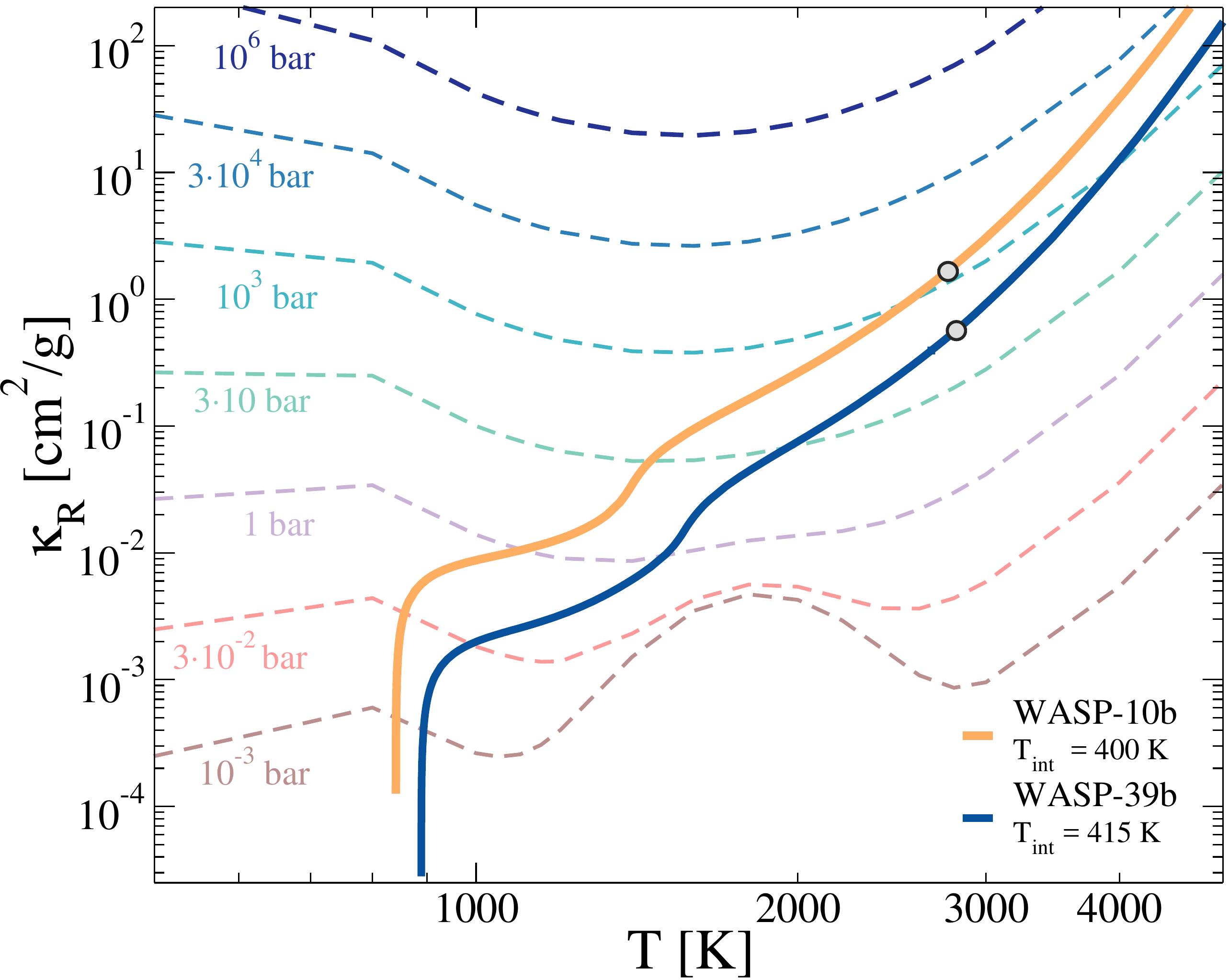}
\caption{{Fit   of    \citet{Valencia2013} to the Rosseland mean opacities $\kappa_{\rm R}$   of    \citet{Freedman2008} 
along isobars (dashed) and $\kappa_R$ along the present $P$--$T$ profiles for WASP-10b (solid orange) and WASP-39b 
(solid blue). The transition from the atmosphere to the interior is marked by grey circles. } }
\label{fig:kappa}
\end{figure}

The results of these fits are shown in Figure~\ref{fig:cloudfreePT}. In Figure~\ref{fig:kappa}, we show the Rosseland mean
opacities along the $P$--$T$ profiles for the present planets using the fit formula of  \citet{Valencia2013} to the 
tabulated values of \citet{Freedman2008}. We conclude that our obtained long-wave opacity values 
$\kappa_{\text{L,0}}\sim 0.01$ cgs are appropriate mean Rosseland mean opacities in the radiative atmospheres of both planets. 

%%%%%%%%%%%%%%%%%%%%%%%%
\subsection{Cloud Model}

The model of~\citet{Heng2012} for a purely absorbing cloud provides a simple toy model approach that reduces the complexity 
of the problem to few parameters {while including the important greenhouse effect of clouds.}  
The cloud opacity can be assumed to take the shape 
\begin{equation}\label{eq:kappac}
\kappa_c (P) = \kappa_{c_0} \exp\left[ - \Delta_c \left(1 - P/P_c\right)^2\right] \quad.
\end{equation}

\begin{figure}[H]
\centering
\includegraphics[width=12 cm]{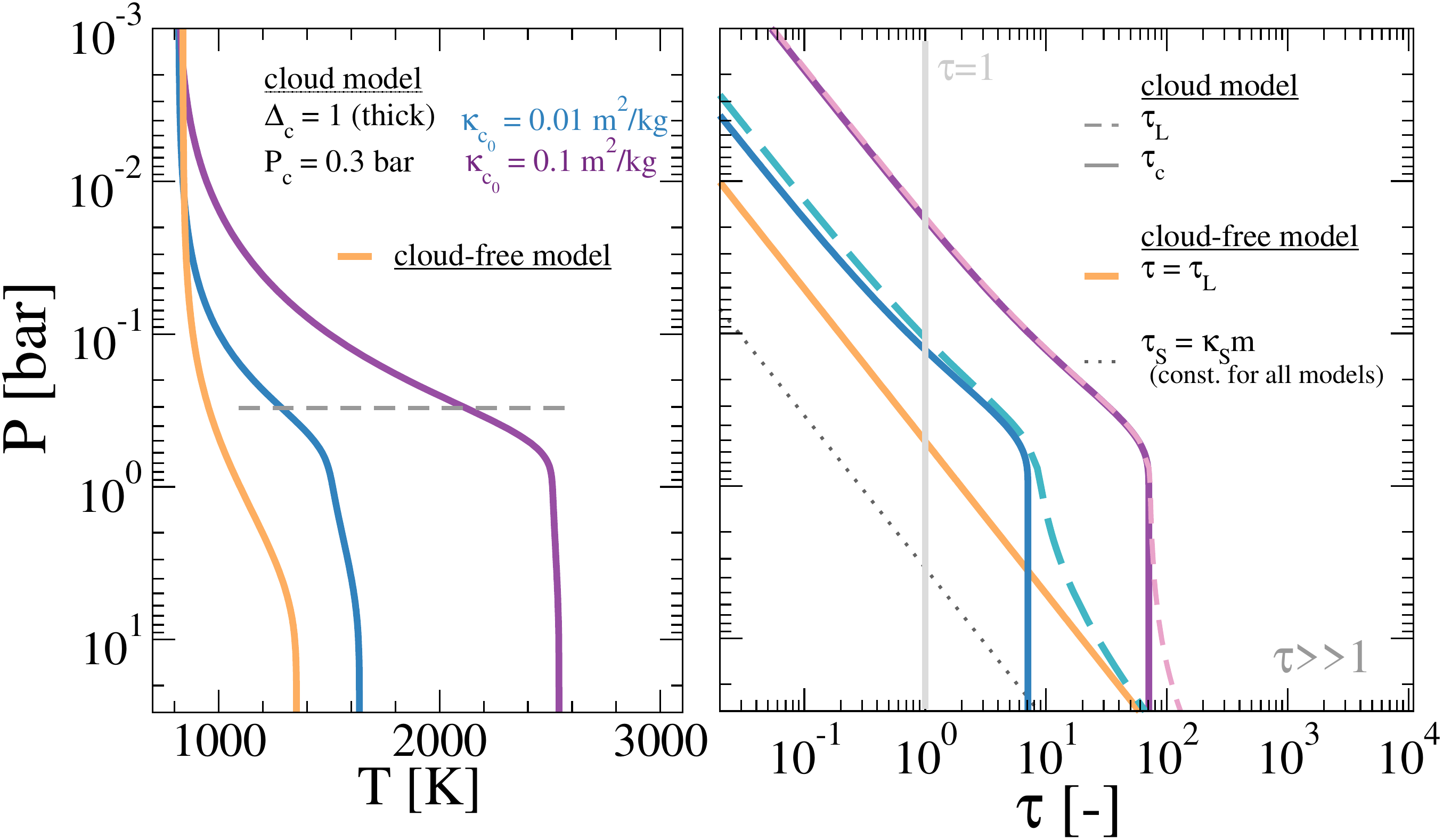}
\caption{
Influence of two cloud decks (optically thin, blue; optically thick, violet, cloud-free, orange) 
on the $P$--$T$ relation of WASP-10b for $T_{\rm int}=0\:$K (left) and $P$--$\tau$-relation (right) {for optical depth 
$\tau_L$ (dashed) and cloud optical depth $\tau_c$ (solid)}, which contributes to $\tau_L$ (cf.~Equation~(49)
in~\cite{Heng2012}). The~cloud decks are located at 0.3 bar and are of vertical extension $\Delta_c=1$. }
\label{fig:wasp10b_pTtau}
\end{figure}

\begin{figure}[H]
\centering
\includegraphics[width=12 cm]{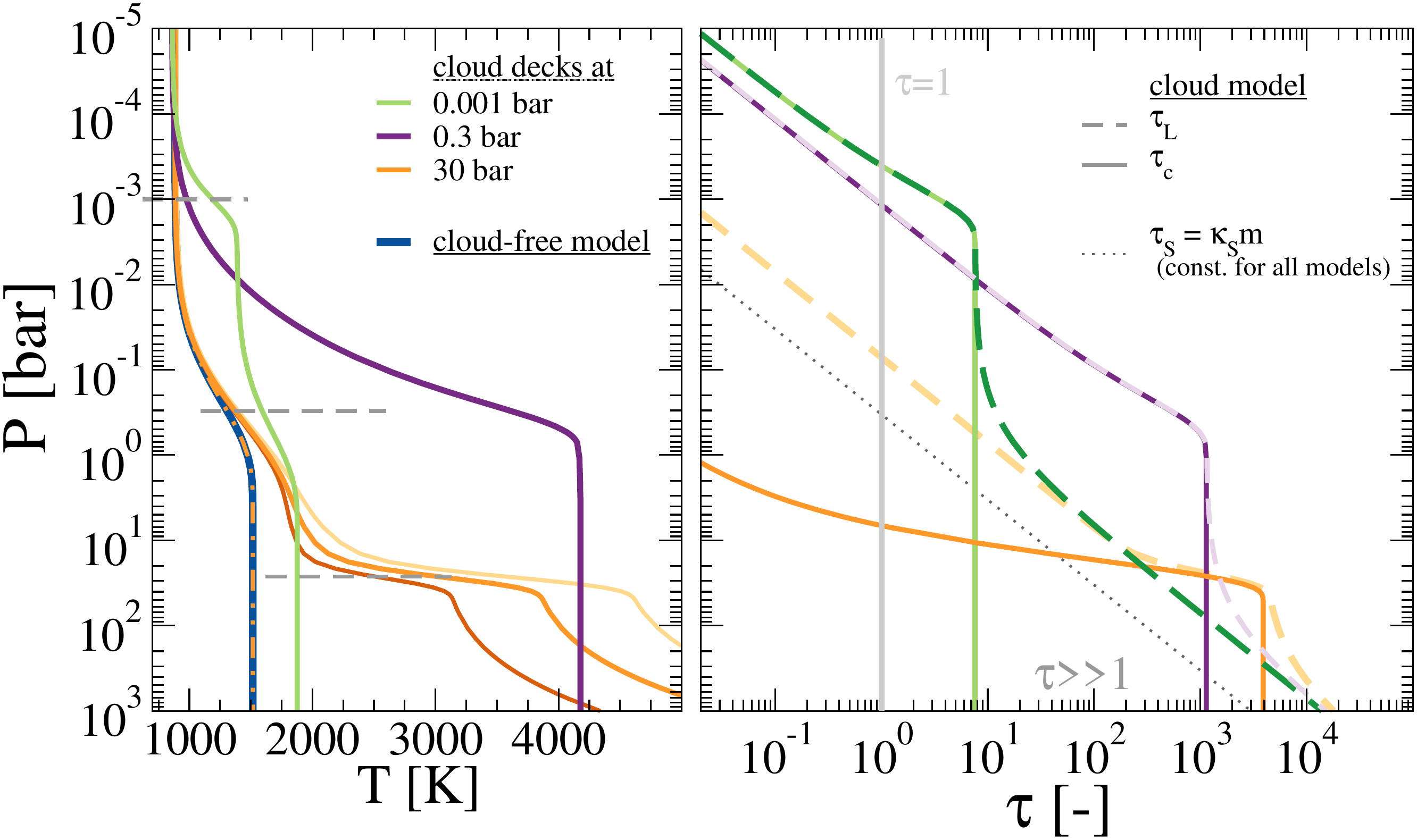}
\caption{
Similar to Figure~\ref{fig:wasp10b_pTtau} but for WASP-39b and three possible cloud locations at 0.001 (green), 
0.3 (purple) and 30 bar (orange) as well as the cloud-free atmosphere (blue, only left panel). {For the cloud deck at 30 bar, we show the resulting $P$--$T$ profiles for $T_{\rm int}=300, 400, 500$ K (from darker to lighter orange) as well as the isotherm (dash-dotted, orange). }}
\label{fig:wasp39b_pTtau}
\end{figure}

The cloud opacity depends on the normalization factor $\kappa_{\rm{c_0}}$, the location of the cloud deck $P_c$, 
and the cloud deck thickness parameter $\Delta_c$, where small $\Delta_c$ values yield vertically extended cloud decks 
while large $\Delta_c$ values lead to thin cloud decks. 
{By construction, the cloud opacity adopts 
a Gaussian shape. The cloud optical depth $\tau_c$ adds to the longwave optical depth $\tau_L$. This is illustrated in Figures~\ref{fig:wasp10b_pTtau} and \ref{fig:wasp39b_pTtau} 
for WASP-10b and WASP-39b, respectively, for cloud parameters considered in this work. The~cloud normalization opacity $\kappa_{c_0}$ was adjusted to reach optical depth values $\tau_{\rm L}$ as in~\cite{Lines2018, Marley2000a}.}
In this cloud model, cloud decks lead to warming of the atmosphere above the cloud, although high
above the cloud deck the effect may reverse and lead to cooling (not shown in Figures~\ref{fig:wasp10b_pTtau} 
and~\ref{fig:wasp39b_pTtau}). 

Nevertheless, the enhancement of opacity and optical depth in a limited region of the atmosphere leads to
a strong heating of the deep atmosphere (left panel) and the typical isothermal region of temperature $T_{\rm iso}$, 
which is most clearly seen for $T_{\rm int}=0$, is shifted toward higher $T_{\rm iso}$ values. For $T_{\rm int}=0$, 
the isothermal region extends all the way down to the center of the planet. Of interest to this study is the 
question of how much the warming effect of the clouds affects the deep interior of planets of finite intrinsic heat fluxes
($T_{\rm int}>0$), and how much this warming effect affects our inferred heavy element abundances.

\unskip

To address this question, we investigate six different possible cloud layers for WASP-10b and three for WASP-39b. 
They are selected based on the condensation curves of typical cloud species~\cite{Lodders2006, Morley2015, 
Kataria2016} shown in Figure~\ref{fig:cloudfreePT}. Possible cloud forming species and their approximate 
intersection pressures are listed in Table \ref{tab:clouddeckparameter}. We take those pressures as the cloud deck
mean location $P_c$ in Equation~(\ref{eq:kappac}). We consider optically thick ($\kappa_{c_0} \sim 10\:\kappa_{\text{L,0}}$,
$\tau_c>1$, see Figures~\ref{fig:wasp10b_pTtau} and \ref{fig:wasp39b_pTtau}) and optically very thick 
($\kappa_{c_0}\sim 100\: \kappa_{\text{L,0}}$, $\tau_c>10$) cloud decks. However, for simplicity, we label them
\emph{optically thin} and \emph{optically thick}, respectively. {The vertical extension is set to $\Delta_c=1$ where possible 
in order to allow for a non-zero ($\Delta_c$ sufficiently large) but not tremendously too strong ($\Delta_c$ sufficiently small) effect.} In the real planet, several cloud decks may be present simultaneously and they may be patchy, 
while, in this model, only one permanent cloud deck is considered and assumed to be uniform.

\begin{table}[H]
\caption{Cloud deck parameters considered in this work.}
\centering
\begin{tabular}{ccccc}
\toprule
			&\textbf{Cloud Species}		& \boldmath{$P_c$} \textbf{[bar]}	& \boldmath{$\Delta_{c}$} \textbf{[-]}& \boldmath{$\kappa_{c_0}$} \textbf{[m$^2$/kg]}	\\
\midrule
	&KCl/ZnS		& 0.01		& 1	& 0.01	\\
			&KCl/ZnS		& 0.01		& 1	& 0.1	\\
			&Na$_2$S 	& 0.3		& 1	& 0.01	\\
\textbf{WASP-10b}			&Na$_2$S 	& 0.3		& 1	& 0.1	\\
			&MnS			& 10		& 10	& 0.01	\\
			&MnS			& 10		& 10	& 0.1	\\
\midrule
	&Na$_2$S		& 0.001		& 1	& 0.2\,\\
	\textbf{WASP-39b}		&MnS			& 0.3		& 1	& 0.1\,\\
			&MgSiO$_3$/Cr		& 30		& 10	& 0.01\,\\
\bottomrule
\end{tabular}
\label{tab:clouddeckparameter}
\end{table}

%%%%%%%%%%%%%%%%%%%%%%%%%%%%%%%%%%%%%%%%%%%
\subsection{Atmosphere-Interior Connection}

The transition to the adiabatic interior is made where the local numeric temperature gradient $\nabla_{\rm T,local}$ is larger than the adiabatic gradient $\nabla_{\rm ad}$ taken from the EOS table. 
Further, we see a convective region forming in most cloudy models above the cloud deck. As the starting point for the adiabatic interior we take the lower intersection of $\nabla_{\rm T, local}$ with $\nabla_{\rm ad}$.  {Generally, the boundary  moves to lower pressures with increasing $T_{\rm int}$ and $T_{\rm eq}$~\cite{Thorngren2019a}}.

%%%%%%%%%%%%%%%%%%%%%%%%%%%%%%%%%%%%%%%%%%%%%%%%%%%%%
\subsection{Planetary Evolution}\label{sec:meth_evol}

\begin{figure}[H]
\centering
\includegraphics[width=10 cm]{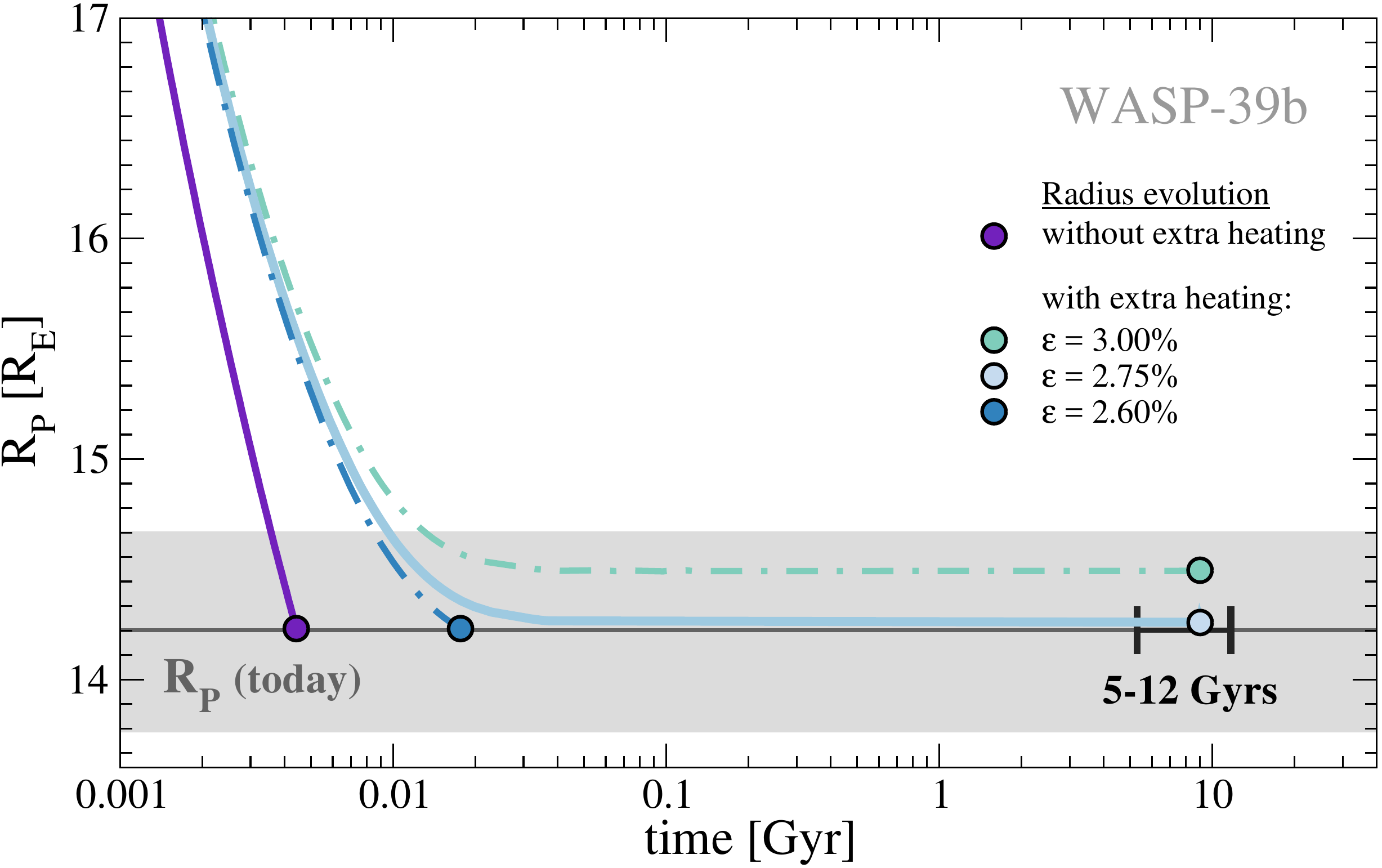}
\caption{Radius evolution for WASP-39b with (bluish) and without (purple) extra heating. The planet stays hot and inflated 
for several Gyrs as the additional heating $\epsilon$ prevents further contraction.   }
\label{fig:WASP39b_evolution}
\end{figure}

To determine the present $T_{\rm int}$ value of a planet, we perform thermal evolution calculations. 
The planets are assumed to be of the same age as the parent star within an uncertainty of a few Myr. Further, 
we assume an orbital location constant in time. Of course, the planets once migrated to their present location,
but this is thought to have happened on a comparably short timescale during the {first 10 Myr~\cite{Alexander2009}}. Integrating the 
{energy balance equation} over time, we obtain the evolution of luminosity $L$ and radius $R_P(\rm t)$
\begin{align}
 L_{\text{eff}} - L_{\text{eq}} = L_{\text{int}} =  L_{\text{sec}}+L_{\text{radio}}+L_{\text{extra}}
\end{align}
with $L_{\text{eq}}= 4\pi R_P^2\sigma_BT_{\text{eq}}^4$ being the absorbed and re-emitted flux. The heat loss from the interior $L_{\text{int}}= 4\pi R_P^2\sigma_BT_{\text{int}}^4$ contains three further components. 
$L_{\text{sec}}=-4\pi R_P^2 \int_{0}^{M} \diff m T(m)\,\frac{\diff s}{\diff t}$ accounts for cooling and contraction 
of the planet, $L_{\text{radio}}$ stands for radiogenic heating, but is of minor importance for H/He-dominated 
gas giants, and $L_{\text{extra}}=\epsilon 4\pi R_P^2\sigma_BT_{\text{eq}}^4$ denotes an extra energy that may be
needed to inflate the planet. In Ref.~\cite{Thorngren2018}, the statistically most likely values of $\epsilon$ as a 
function of irradiation flux are determined for a sample of planets that exclude planets with $M_P<0.5\:M_{\rm Jup}$. 
Here, we need the extra heating term in order to reach the {large age} of the 0.28 $M_{\rm Jup}$ planet WASP-39b. 
Depending on the distribution of heavy elements in the envelope vs. core, we find $\epsilon$ = 2.75--4.00\% compared to 
the majority of hot Jupiters where $\epsilon$ = 1--3\%~\cite{Thorngren2018}. In Figure~\ref{fig:WASP39b_evolution}, we show 
the radius evolution of WASP-39b with and without extra heating. For the young WASP-10b, we do not need 
extra heating to explain its measured radius.

%%%%%%%%%%%%%%%%%%%%%%%%%%%%%%%%%%%%%%%%%%%%%%%%%%%%%%%%%%%%%%%%%%%%%%%%%%%%%%%%%%%%
\section{Results for WASP-10b}\label{sec:resultsWASP10b}

\begin{figure}[H]
\centering
\includegraphics[width=10 cm]{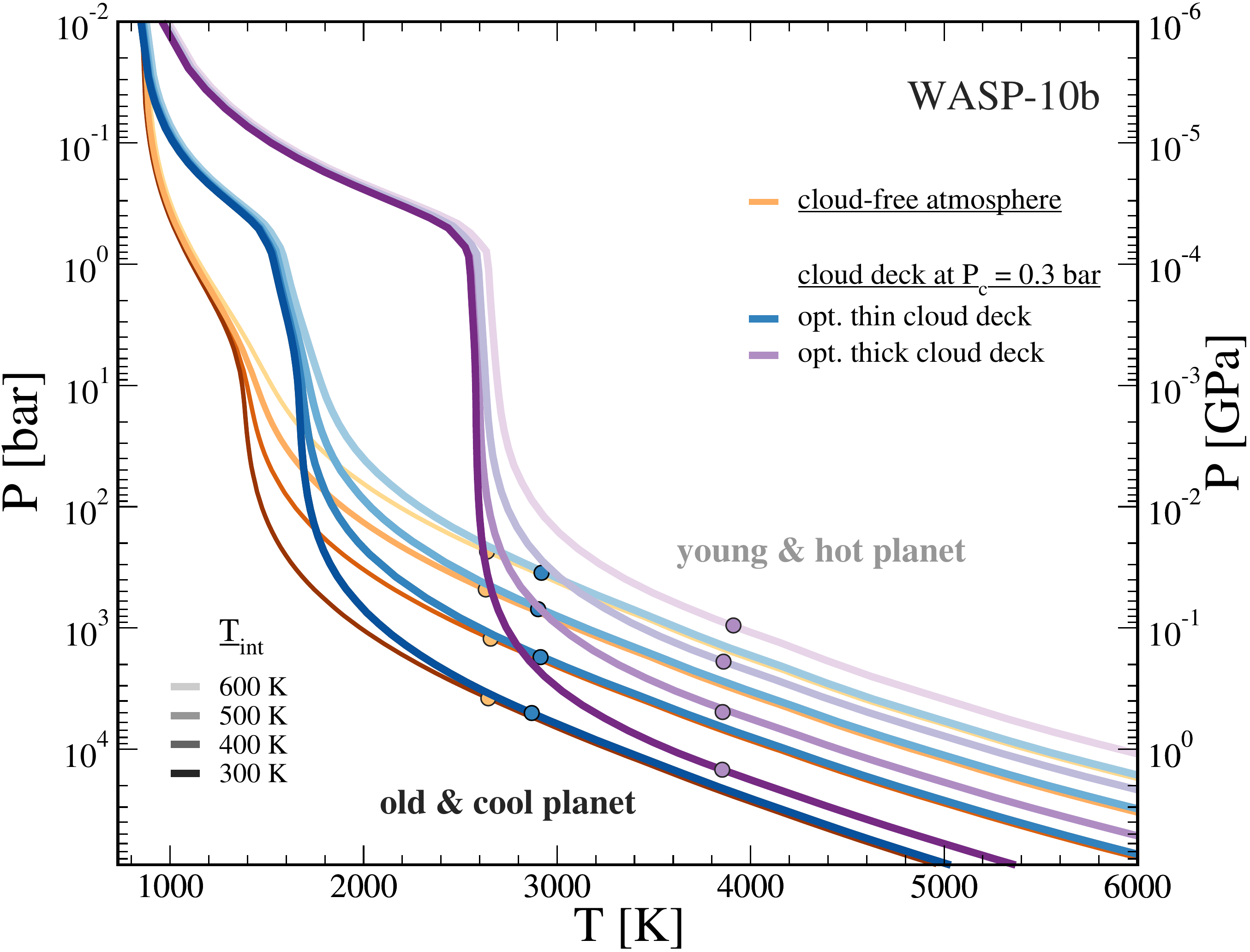}
\caption{Atmospheric $P$--$T$ profiles of WASP-10b for finite $T_{\text{int}}$ values of 300, 400, 500, and 600~K (from bottom to top) and three cloud deck scenarios: optically thin at 0.3 bar (blue), optically thick at 0.3 bar (purple), and  cloud-free  (orange). Circles mark the transition between atmosphere and adiabatic interior.}
\label{fig:WASP10b_PT}
\end{figure}

In Figure~\ref{fig:WASP10b_PT}, we show atmospheric $P$--$T$ profiles for  WASP-10b for finite $T_{\rm int}$ values 
for a cloud deck at $0.3\:$bar and two  different cloud opacities $\kappa_c=0.01$ and 0.1 m$^2$/kg.

As   shown  in Figure~\ref{fig:wasp10b_pTtau} for $T_{\rm int}=0\:$K, clouds can shift the temperature in the isothermal region significantly toward higher values. This is also the case for finite $T_{\rm int}$ values. Figure~\ref{fig:WASP10b_PT} also
shows that clouds shift the onset of the adiabatic interior to deeper regions. Both effects become more pronounced with increasing cloud opacity. {However, we find that the interior adiabat follows 
the adiabat of the cloud-free case of same $T_{\rm int}$ value.}

%%% optically thick vs. thin cloud %%%
The optically thin cloud (blue) in Figure~\ref{fig:WASP10b_PT} shifts $T_{\rm iso}$ by about 400 K from $\sim$1400 K to 
$\sim$1800 K. Under these conditions, the initially assumed Na$_2$S molecules would no longer condense while silicate clouds (Mg$_2$SiO$_3$ and Mg$_2$SiO$_4$) might form in present WASP-10b. In this sense, we consider the optically thin cloud 
at 0.3 bar a {more likely option} for WASP-10b. On the other hand, the assumption of the optically thick cloud 
at 0.3 bar (purple) clearly shifts $T_{\rm iso}$ far beyond any temperature regime where heavy elements might 
condense out.
Similar reasoning applies to the   four other cloud cases considered for WASP-10b. Our optically thick clouds cause too strong 
heating, evaporating any clouds, while in the atmosphere heated by the optically thin clouds condensible species could still
condense out. {This is the picture that emerges if using condensation curves for solar-metallicity atmospheres.} Despite the apparent inconsistencies with the optically thick clouds, we keep them in the loop. This allows 
us to place an upper limit on the quantitative influence of assumed long-wave absorbers on the inferred metallicity.

\begin{figure}[H]
\centering
\includegraphics[width=12 cm]{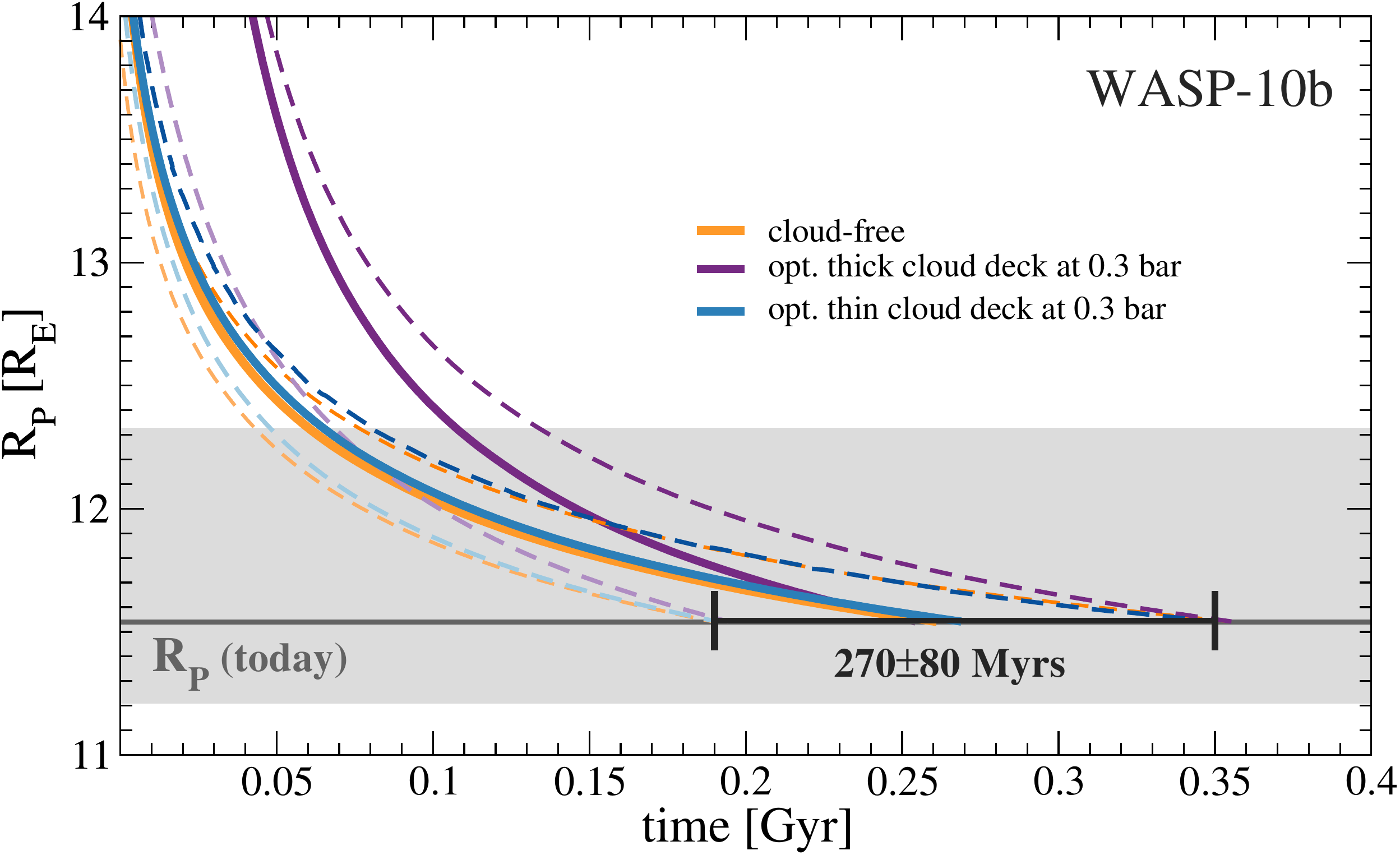}
\caption{{Radius evolution of WASP-10b for the cloud-free atmosphere (orange) as well as for optical thick (purple) and optical thin (blue) clouds decks at $P_c=0.3$ bar. Solid lines yield a cooling time of 270 Myr while dashed lines within the $1\sigma$
uncertainty of the age, e.g., light dashed lines then describe models reaching the lower limit of the age constraint of 190 Myr.}}
\label{fig:WASP10bevolution}
\end{figure}

%%%%%%%%%%%%%%%%%%%%%%%%%%%%%%%%%%%%
We proceed with the case of the cloud at $0.3$ bar and show the radius evolution in Figure~\ref{fig:WASP10bevolution}.  

\begin{figure}[H]
\centering
\includegraphics[width=10 cm]{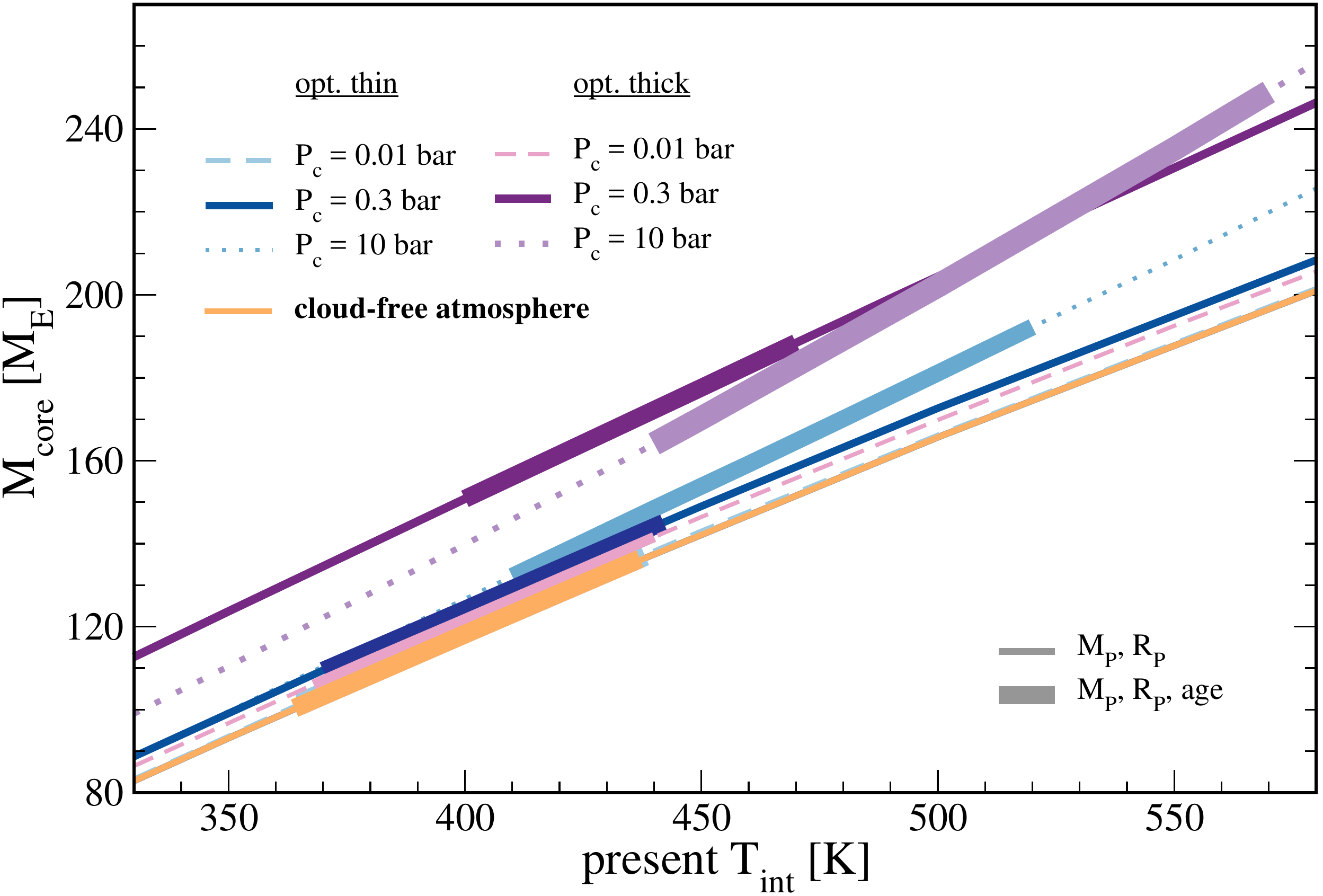}
\caption{$M_{\text{core}}$-$T_{\text{int}}$ relation for WASP-10b assuming cloud-free atmosphere (orange) and six different 
cloud decks in the atmosphere. Thin lines indicate the results obtained by $M_P$, $R_P$, thick bars highlight the
solutions that also satisfy the age constraint.}
\label{fig:WASP10b:TintMcore}
\end{figure}

Cooling times in agreement with the known age of the system can easily be obtained for all considered cloud models. 
Lower assumed $T_{\text{int}}$ values for the present planet lead to longer cooling times. We find that optically thick clouds with their
strong heating effect slow down the heat loss from the interior, leading to higher $T_{\text{int}}$ values. They also slow down the
contraction of the planet. To obtain a radius for the present planet in agreement with the known age and radius, 
the planet with optically very thick clouds must harbor a larger amount of heavy elements. 
That leads to the link between $M_{\text{core}}$ in representation of planetary bulk heavy element mass $Z_P$ and 
$T_{\text{int}}(t_0)$ shown in Figure~\ref{fig:WASP10b:TintMcore}.
%%% Figure 8 %%%
Thick lines in Figure~\ref{fig:WASP10b:TintMcore} show the models matching $R_P$, $M_P$ and the error range of the age of the system. 

The higher is $T_{\rm iso}$, rising with the optical thickness and $P_c$, the larger is the core mass. The {more likely option of the optically thin cloud deck at $0.3$ bar (solid dark blue) leads to a 10\% higher core mass compared to the cloud-free model (orange).
 For optically thin clouds high in the atmosphere,  the heating effect on the atmosphere is lower and
the influence on inferred metallicity is negligible (the~blue-dashed curve in Figure~\ref{fig:WASP10b:TintMcore} coincides with 
the orange curve). For} the optically thin clouds deep in the atmosphere the heating
effect is strong and therefore we had to make the cloud more tenuous by increasing $\Delta_c$ instead. The maximum enhancement
in inferred heavy element abundance is about 10\% and well represented by the medium-height cloud at $0.3$ bar.
For optically thick clouds, which are not {likely options}, we obtain a maximum increase in inferred heavy element 
content of up to 100\%.  

Thorngren   and   Fortney (2019) find for WASP-10b $Z_{\rm P}=0.12\pm0.02$, using $M_P=3.15\,M_{\rm Jup}$ and 
$R_P=1.08\,R_{ \rm Jup}$, in agreement with our results for the cloud-free model, where we obtain $Z_{\rm P}= 0.13$
at $T_{\rm int}=400$~K.

%%%%%%%%%%%%%%%%%%%%%%%%%%%%%%%%%%%%%%%%%%
%%%%%%%%%%%%%%%%%%%%%%%%%%%%%%%%%%%%%%%%%%
%%%%%%%%%%%%%%%%%%%%%%%%%%%%%%%%%%%%%%%%%%

\section{Results for WASP-39b}
\label{sec:resultsWASP39b}
\vspace{-8pt}

\subsection{Cloud Height}

\begin{figure}[H]
\centering
\includegraphics[width=10 cm]{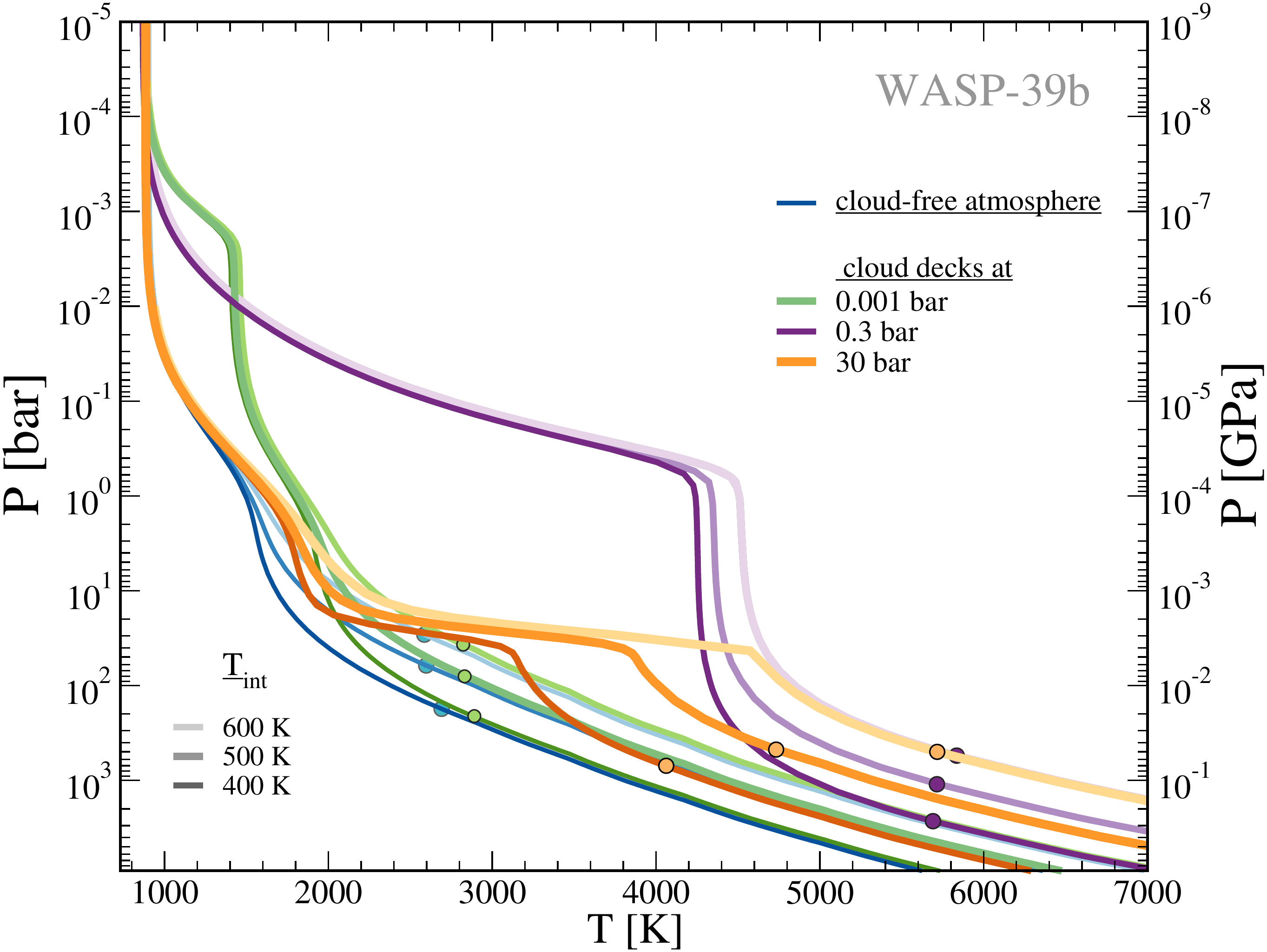}
\caption{Atmospheric $P$--$T$ profiles of WASP-39b for finite $T_{\text{int}}$ values of 400, 500, and 600~K for three decks 
located at 0.001 (green), 0.3 (purple), and 30 (orange) bar as well as the cloud-free atmosphere (blue). Circles mark the 
transition between atmosphere and adiabatic interior.}
\label{fig:WASP39b_PT}
\end{figure}

In Figure~\ref{fig:WASP39b_PT}, we show the atmospheric $P$--$T$ profiles for  WASP-39b with and without cloud decks. 
We find that the high cloud deck at 0.001 bar would heat the upper atmosphere so much that only silicates could condense out 
at such low pressures.

The analysis in Ref.~\cite{Wakeford2018} indicates the presence of clouds on only one side of the limb while a clear sky 
on the other. In their 3D global circulation models, a high-metallicity atmosphere was clearly required to explain the 
spectra while optically thick, {uniform} clouds would not much influence the fit. Thus, the observations do not well constrain the presence of clouds, in particular in the deep atmosphere below  $\sim$0.1 bar or deeper. {We proceed with 
the cloud deck at 30 bar.
According to Figures~\ref{fig:cloudfreePT} and~\ref{fig:WASP39b_PT}, this cloud deck could be a more likely solution
for the 10--30 bar region while the heating of the deeper troposphere for the deep-seated cloud at 30 bar is very strong.}
At 3000--4000~K, condensible species will not condense out. On the other hand, a uniform silicate cloud layer at 30 bar may
impose a compositional gradient, which itself may inhibit convection unless the super-adiabaticity becomes sufficiently
strong. As a result, the temperature gradient needed to transport the internal heat outward must be larger than in the adiabatic case without cloud. For the solar system giant planets, this effect may amount up to several 100 Kelvins~\cite{Leconte2017}. Therefore, we consider the deep, optically thin cloud at 30 bar a possible option for WASP-39b. {We caution that a number of further effects may lead to a more complex picture than drawn here. Condensation of heavier species decreases the mean molecular weight of the surrounding medium and condensates may decouple from the gas phase, affecting the density difference between vertically moving 
parcels and the background state and thus the possible stability. Moreover, since the Rosseland mean opacity depends on metallicity~\cite{Freedman2008}, redistribution of condensible species by condensation also influences the radiative gradient of the background state. Leconte et al. (2017)~\cite{Leconte2017} also found that possible 
stability requires a sufficiently high mixing ratio of condensible species. Whether sufficient conditions for stability are satisfied in the atmospheres of the hot Jupiters remains to be investigated.}

%%%%%%%%%%%%%%%%%%%%%%%%%%%%%%%%%%%
\subsection{Metallicity}
%%% metallicity and observations%%%

WASP-39b is an interesting planet because of its observationally determined atmospheric water abundance.  
Recently, Wakeford et al. (2018) completed the existing transmission spectrum data in the optical obtained 
with HST STIS~\cite{Sing2016} and VLT FORS2~\cite{Nikolov2016} and in the infrared obtained with \emph{Spitzer} IRAC~
\cite{Sing2016} by adding spectral data in the near infrared using the HST WFC3 camera. The clearly detected water 
absorption features allowed them to retrieve the atmospheric metallicity, temperature, and cloudiness of the 
observationally accessible part of the atmosphere amongst other parameters. 
Combined likelihood analysis of their isothermal equilibrium model yielded a high-metallicity atmosphere
of $\sim$$151^{+48}_{-46}\times$ solar abundances, though their free-chemistry model yielded a lower metallicity of 
$\sim$$117^{+14}_{-30}\times$ solar abundances.

The high metallicity of 100--200$\times$ solar corresponds to a heavy element mass fraction $Z_{\rm env}\sim
0.25$--0.75 (see Table \ref{tab:ZpZenv}). Cloud-free structure model of WASP-39b yield a maximum $Z_p$ value of 0.25
\cite{Thorngren2019}, where $Z_p \geq Z_{\rm env}$ due to the possible presence of a core.
 
\begin{figure}[H]
\centering
\includegraphics[width=12 cm]{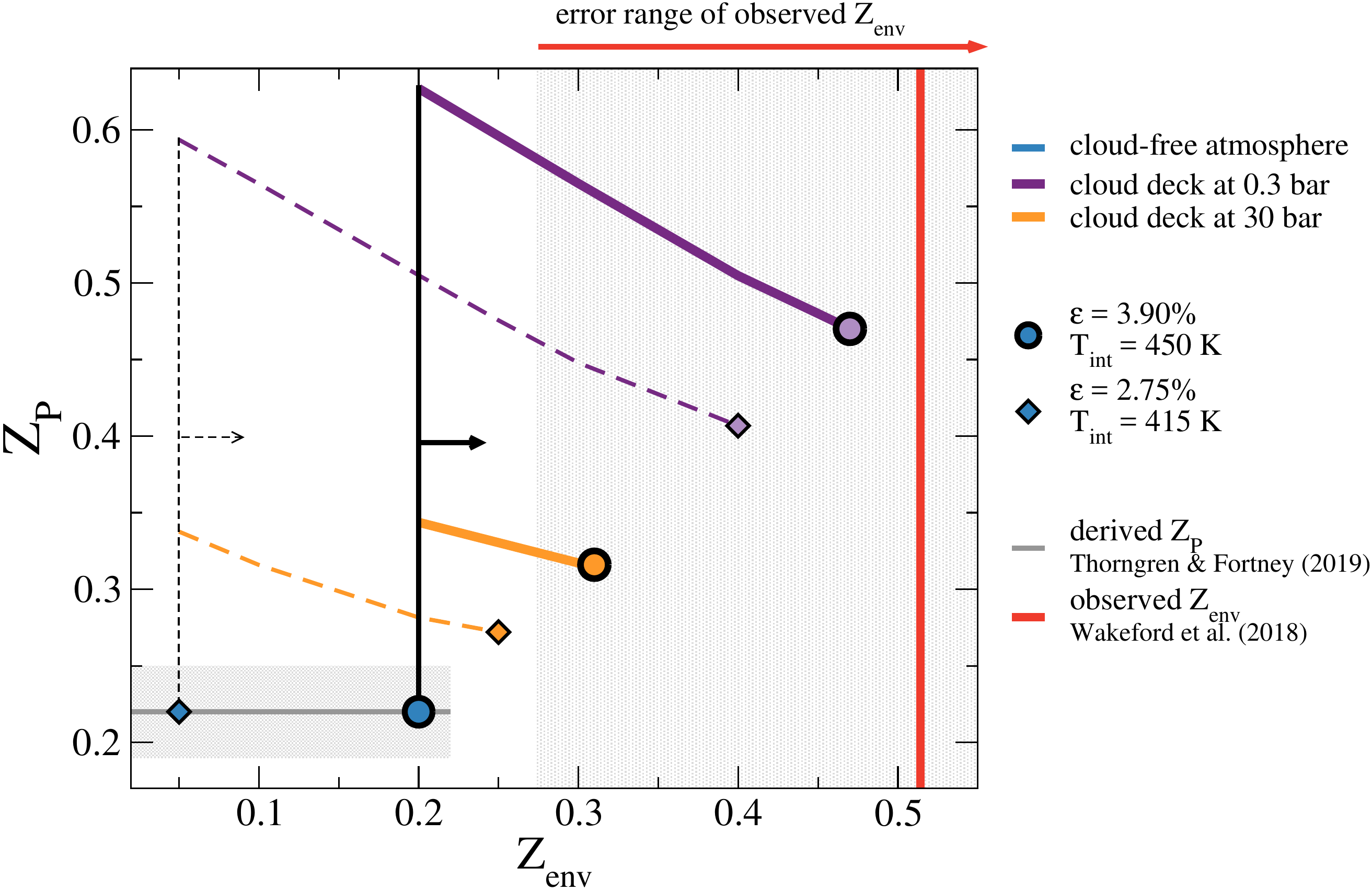}
\caption{Range of envelope metallicity $Z_ {\rm env}$ due to different atmosphere models without clouds (blue), with 
optically thick cloud deck at $0.3\:$bar (purple), and optically thin at $30\:$bar (favored case, orange). 
Circles/diamonds indicate the maximum $Z_ {\rm env}$ value for a fully-mixed planet. The solid and dashed colored 
lines indicate models with different $T_{\rm int}$/$\epsilon$-values obtained for the cloud-free models. 
The arrows indicate the increase of $Z_ {\rm env}$ when clouds are switched on.  }
\label{fig:WASP39bresults}
\end{figure}

First, we require our cloud-free models to have $Z_P=0.22\pm 0.03$ as found in Ref.~\cite{Thorngren2019} for cloud-free models.
Because this planet seems to be inflated (see Section~\ref{sec:meth_evol}), we account for extra heating $\epsilon>0$.
For $Z_{\text{env}}=0.05$, we find $\epsilon\approx2.75\%$, whereas for $Z_{\text{env}}=0.2$ we find $\epsilon\approx3.90\%$.
These $\epsilon$ values are then used also for the models with clouds. From our experience with the models for WASP-10b, 
where optically thin clouds have a minor effect on $T_{\rm int}$, we also use the same range of $T_{\rm int}$ values 
as found for the cloud-free case, so that no additional evolution calculations are necessary. 
Figure~\ref{fig:WASP39bresults} shows the results on $Z_{\rm env}$ and $Z_p$.
Even for the extreme case of the optically thick cloud, high atmospheric metallicities of $Z_{\rm env} \sim 0.5$ as observationally derived can barely be reached. Interestingly, however, for the optically thin deep cloud the enhancement in inferred metallicity amounts to about 50\%, which allows us to obtain solutions just \emph{within} the observational uncertainty   of    \citet{Wakeford2018}. 
A summary of the metallicities for WASP-39b is given in Table \ref{tab:ZpZenv}.

With our favored cloud model for WASP-39b, the optically thin deep cloud, we obtain a maximum envelope metallicity of 0.3,
which is 50\% higher than our value in the cloud-free case. However, the maximum $Z_{\rm env}$ value still falls short of
the observed value. Our results therefore confirm the conclusion of Thorngren   and   Fortney (2019) of additional 
sources of uncertainty relevant to WASP-39b. One source of uncertainty is
the EOS. While the H/He-EOS was found to induce an uncertainty of a few percent only for massive hot Jupiters
and brown dwarfs, wherein matter is largely degenerate \citep{Becker2014}, this effect might be stronger for warm, lower-mass 
planets where temperature effect on the $P$--$\rho$ relation can be stronger. The composition of heavy elements
matters as well. Icy cores typically have a 50\% higher mass than rocky cores if otherwise the same modeling procedure is applied.
\unskip
\begin{table}[H]
\caption{Constraints on atmospheric metallicity of WASP-39b.}
\label{tab:ZpZenv}
\centering
\begin{tabular}{cccc}
\toprule 
\textbf{}	& \textbf{WASP39-b} \\
\midrule
Wakeford et al.	(iso. eq.)	& [M/H]=$151^{+48}_{-46} \,\times$ solar		& $Z_{\rm env}=0.514^{+0.25}_{-0.24}$ &\\
Wakeford et al.	(free-chem.)	& [M/H]=$117^{+14}_{-30}\, \times$ solar		& $Z_{\rm env}=0.45^{+0.09}_{-0.17}$ &\\
Thorngren   and   Fortney 	& Z:H$_{\text{P}}=40.51\pm8.3\, \times$ solar	&$Z_P=0.22\pm0.03$ (=$Z_{\rm env}$, fully mixed)\\
This work, cloud-free 		& 						&$Z_{\rm env,max}=0.2$  & \\
This work, $0.3\:$bar cloud deck& 						&$Z_{\rm env,max}=0.47$ & \\
This work, $30\:$bar cloud deck	& 						&$Z_{\rm env,max}=0.31$ & \\
\bottomrule
\end{tabular}\\
\begin{tabular}{@{}c@{}} 
\multicolumn{1}{p{\textwidth -.88in}}{\footnotesize \textbf{Notes.} For conversion of [M/H] to Z, we use Equation~(3) in~\cite{Thorngren2019} with water as heavy element. The ratio Z:H$_P$ is the atmosphere abundance for a fully mixed planet, as derived derived from interior models in~\cite{Thorngren2019}, Equation~(3).}
\end{tabular}
\end{table}

Since WASP-39b is likely to be heavy element-rich, it could also be that the heavy elements are not homogeneously distributed
but that their abundance increases with depth. Even slight compositional gradients can suppress convection and delay cooling.
This may be the case in exoplanets~\cite{Chabrier2007} and in Saturn itself~\cite{Leconte2013}.

\section{Comparison to Self-Consistent Cloud Models}
\label{sec:ComparisonAtmosphere}

\begin{figure}[H]
\centering
\includegraphics[width=14 cm]{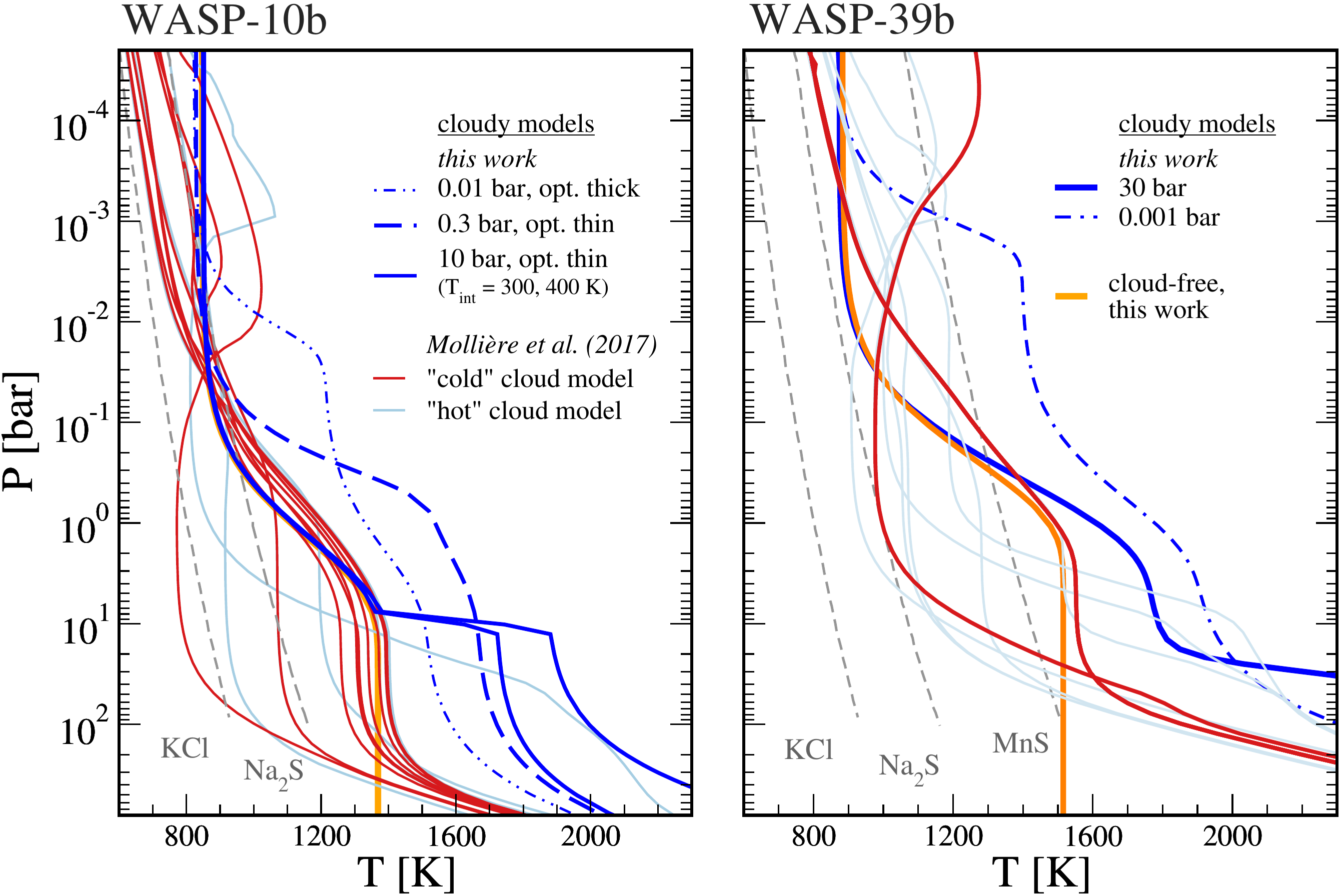}
\caption{Atmospheric temperature structures for WASP-10b ({left}) and WASP-39b ({right}). Our result for the clear atmosphere is shown in orange for $T_{\rm int}=0$ K as well some of our cloudy solutions in dark blue. The bunch of profiles in red (\textit{cold models}) and light blue (\textit{hot models}) are from Molli\`{e}re et al. (2017) for different cloud model parameter, see text. The condensation curves assume solar elemental abundance.}
\label{fig:WASP10b39bMolliere}
\end{figure}

Clouds will not only be important at infrared wavelengths, but they can also contribute to absorption and scattering of irradiation at short wavelengths. This is neglected in the cloud model we use.
Out of the codes capable of calculating the structure of self-luminous and/or irradiated planets  (e.g.,  ~\cite{Malik2019, Allard2001, Hauschildt1999}), we here compare the ad-hoc approach of Heng et al. (2012)~\cite{Heng2012} to the self-consistent atmosphere models with clouds of Molli\`{e}re et al. (2017)~\cite{Molliere2017}, who used the \textit{petitCODE}~\cite{Molliere2015}. Within this code, models with clouds and different metallicities have been calculated specifically for WASP-10b and WASP-39b. 
That code calculates radiative-convective equilibrium atmospheric structures and spectra of extrasolar planets  self-consistently, assuming chemical equilibrium. The radiative transfer model implements absorption, emission and scattering. It implements the Ackerman   and   Marley (2001)~\cite{Ackerman2001} cloud model for clouds composed of MgAl$_2$O$_4$, Mg$_2$SiO$_4$, Fe, KCl and Na$_2$S. Particle opacities are calculated using Mie theory (assumption of spherical, homogeneous grains) or the distribution of hollow spheres (approximating irregularly shaped dust aggregates). 
For both planets, WASP-10b and WASP-39b, we plot the clear and cloudy solutions of Molli\`{e}re et al. (2017) in comparison to our clear and cloudy atmosphere models for $T_{\rm int} = 400$ K in Figure~\ref{fig:WASP10b39bMolliere}. 
 Molli\`{e}re et al. (2017) used cloud models which differ in the assumptions of the grain shape, the standard settling parameter $f_{sed}$ from the Ackerman   and   Marley model, the maximum cloud mass fraction, the width of the cloud particle size distribution as well the inclusion of iron clouds 
 (see Table 2 in~\cite{Molliere2017}). The different model assumptions result in different atmospheric structures. For temperate giant planets, such as WASP-10b and WASP-39b, they investigated cold cloud models as well, where only Na$_2$S and KCl are considered as possible cloud species (Figure~\ref{fig:WASP10b39bMolliere}, red) as for this temperature regime higher temperature condensates may not mix up from their deep cloud deck locations. 

For both planets, the cloudy atmospheric structures from Molli\`{e}re et al. (2017) lead to both cooler and hotter isotherms. Their favored \textit{cold} cloud models, only using Na$_2$S and KCl as cloud species, lead to cooler isotherm for all different cloudy model parameters compared to the clear atmosphere in orange. In contrast, in this work, the fit parameter of the double-gray atmosphere and the added cloud opacity lead to a warmer atmosphere beneath the cloud deck for all of our assumed cloud decks in the atmosphere. For WASP-10b, there is only one model (dashed light blue) that yields a hotter isotherm, whereas for WASP-39b there are three cloudy models that yield a hotter isotherm compared to the cloud-free case. 
This comparison suggest that our favored 0.3 bar cloud model for WASP-10b may be supported by the \textit{hot} cloud model of Molli\`{e}re et al. (2017), down to significant depths of $\sim$1~kbar, while for WASP-39b our favored 30 bar cloud model is supported to $\sim$100~bar and thus may overestimate the here obtained influence on the $Z_{\rm env}$ of WASP-39b.

%%%%%%%%%%%%%%%%%%%%%%%%%%%%%%%%%%%%%%%%%%
\section{Conclusions}
\label{sec:conclusions}

We   performed coupled atmosphere, interior, and evolution calculations for the two giant planets WASP-10b and WASP-39b
and investigated the effect of   additional absorbers, which we call   cloud decks, of the inferred metallicity. 
We assumed cloud optical thicknesses of $\tau_c \approx$ 1--10 (named optically thin) and $\tau_c\approx$ 1--100 
(named optically thick), as well as different cloud heights in the atmosphere corresponding to certain condensible species. 
The clouds decks are purely absorbing and based on the model of Heng et al. (2012)~\cite{Heng2012}. 
Our major findings are as follows:

\begin{enumerate}[leftmargin=21pt,labelsep=7pt]
\item[(I)] Through their additional infrared opacity, {these cloud decks tend to warm the atmosphere beneath}. This leads to a more or less pronounced enhancement in inferred heavy element abundance.

\item[(II)] {For the optically thicker cloud decks,} the heating is too strong so that condensible species would no  longer condense out. This puts
an upper limit on the enhancement in metallicity of 100\% on both planets. 

\item[(III)]  For optically thin clouds, the heating of the atmosphere can be sufficiently small so that condensible 
species can condense out. In this case, we find an increase of the core mass of up to 10\% for WASP-10b. 

\item[(IV)]  For WASP-39b we find a maximum atmospheric metallicity of $Z_{\rm env}=0.31$ if we assume a deep cloud at 30 bars 
in the troposphere that in addition would lead to inhibited convection. Even in this {favored} case, the possible 
envelope metallicity is still near the lower limit of the observationally inferred value. 
{Further effects that lead to a heating of the planet are clearly required. Since the heating efficiency is empirically not yet constrained, as it is for hot Jupiters~\cite{Thorngren2018}, $\epsilon$ > 3\% is not excluded for such planets. Such high values would help to bring the observationally inferred $Z$ in agreement with interior-model inferred $Z$.}
\end{enumerate}

Due to the complexity of modeling clouds in a realistic manner, we applied a simple cloud model that is a crude representation 
of real cloud decks. The predictive power of that model stands and falls with the assumed cloud opacity, cloud height, and cloud thickness, which are poorly known parameters. {Another important point is how to couple this cloud model to the atmospheric structure self-consistently.} Nevertheless, this work suggests that deep cloud decks below 
the pressure level accessible to transmission spectra observations may influence the temperature structure in the 
atmosphere and the inferred metallicity to some extent. A more sophisticated approach is desired.
\vspace{6pt}

%%%%%%%%%%%%%%%%%%%%%%%%%%%%%%%%%%%%%%%%%%
\authorcontributions{
A.J.P.   developed large parts of the computer code used to perform the model computations, obtained the results presented here, and prepared the  figures.
A.J.P. and N.N.   wrote the paper, and R.R. approved it. N.N. and R.R.   designed the project. All authors  discussed and contributed to the methodology.}

%%%%%%%%%%%%%%%%%%%%%%%%%%%%%%%%%%%%%%%%%%
\funding{A.J.P. and R.R. thank the DFG for support within the SPP 1992.} 

%%%%%%%%%%%%%%%%%%%%%%%%%%%%%%%%%%%%%%%%%%
\acknowledgments{{We thank the referees for helping to improve this manuscript.} We thank R.~Neuhäuser and G.~Maciejewski for discussions on the
observational parameters of WASP-10b. N.N. thanks the Swiss PlanetS Program for hospitality in March 2018. {We thank P. Molli\`{e}re for providing us with data from petitCODE for the atmospheres of WASP-10b and WASP-39b, and thank J.~Fortney and L.~Scheibe for discussions.} }

\conflictsofinterest{The authors declare no conflict of interest.} 
%%%%%%%%%%%%%%%%%%%%%%%%%%%%%%%%%%%%%%%%%%
%% optional
% \appendixtitles{no} %Leave argument "no" if all appendix headings stay EMPTY (then no dot is printed after "Appendix A"). If the appendix sections contain a heading then change the argument to "yes".
% \appendix
% \section{}
% \unskip
% \subsection{}
% The appendix is an optional section that can contain details and data supplemental to the main text. For example, explanations of experimental details that would disrupt the flow of the main text, but nonetheless remain crucial to understanding and reproducing the research shown; figures of replicates for experiments of which representative data is shown in the main text can be added here if brief, or as Supplementary data. Mathematical proofs of results not central to the paper can be added as an appendix.
% 
% \section{}
% All appendix sections must be cited in the main text. In the appendixes, Figures, Tables, etc. should be labeled starting with `A', e.g., Figure A1, Figure A2, etc. 


%=====================================
% References, variant B: external bibliography
%=====================================
% \externalbibliography{yes}
\reftitle{References}

%%%%%%%%%%%%%%%%%%%%%%%%%%%%%%%%%%%%%%%%%%
%% optional
% \sampleavailability{Samples of the compounds ...... are available from the authors.}

%% for journal Sci
%\reviewreports{\\
%Reviewer 1 comments and authors’ response\\
%Reviewer 2 comments and authors’ response\\
%Reviewer 3 comments and authors’ response
%}

%%%%%%%%%%%%%%%%%%%%%%%%%%%%%%%%%%%%%%%%%%

\begin{thebibliography}{999}

\bibitem[Venturini    {et~al.}(2016)Venturini, Alibert, and
  Benz]{Venturini2016}
Venturini, J.; Alibert, Y.; Benz, W.
\newblock {Planet formation with envelope enrichment: New insights on planetary
  diversity}.
\newblock {\em A{\&}A} {\bf 2016}, {\em 596},~A90.

\bibitem[Wakeford    {et~al.}(2018)Wakeford, Sing, Deming, Lewis, Goyal,
  Wilson, Barstow, Kataria, Drummond, Evans, Carter, Nikolov, Knutson,
  Ballester, and Mandell]{Wakeford2018}
Wakeford, H.R.; Sing, D.K.; Deming, D.; Lewis, N.K.; Goyal, J.; Wilson, T.J.;
  Barstow, J.; Kataria, T.; Drummond, B.; Evans, T.M.; et al.
\newblock {The complete transmission spectrum of WASP-39b with a precise water
  constraint}.
\newblock {\em Astron. J.} {\bf 2018}, {\em 155},~29.

\bibitem[Thorngren and Fortney(2019)]{Thorngren2019}
Thorngren, D.; Fortney, J.J.
\newblock {Connecting giant planet atmosphere and interior modeling:
  constraints on atmospheric metal enrichment}.
\newblock {\em Astrophys. J. Lett.} {\bf 2019}, {\em 874},~6.

\bibitem[Maciejewski    {et~al.}(2011)Maciejewski, Raetz, Nettelmann, Seeliger,
  Adam, Nowak, and Neuh{\"{a}}user]{Maciejewski2011a}
Maciejewski, G.; Raetz, S.; Nettelmann, N.; Seeliger, M.; Adam, C.; Nowak, G.;
  Neuh{\"{a}}user, R.
\newblock {Analysis of new high-precision transit light curves of WASP-10 b:
  starspot occultations, small planetary radius, and high metallicity}.
\newblock {\em A{\&}A} {\bf 2011}, {\em 535},~A7.

\bibitem[Thorngren    {et~al.}(2016)Thorngren, Fortney, Murray-Clay, and
  Lopez]{Thorngren2016}
Thorngren, D.P.; Fortney, J.J.; Murray-Clay, R.A.; Lopez, E.D.
\newblock {The mass-metallicity relation for giant planets}.
\newblock {\em Astrophys. J.} {\bf 2016}, {\em 831},~64.

\bibitem[Morley    {et~al.}(2013)Morley, Fortney, {Kempton M.-R.}, Marley,
  Vissher, and K.]{Morley13gj1214}
Morley, C.V.; Fortney, J.J.; {Kempton}, E.M.-R.; Marley, M.S.; Vissher, C.;
 Zahnle, K.
\newblock {Quantitatively assessing the role of clouds in the transmission
  spectrum of GJ 1214b}.
\newblock {\em Astrophys. J.} {\bf 2013}, {\em 775},~33.

\bibitem[Lines    {et~al.}(2018)Lines, Mayne, Boutle, Manners, Lee, Helling,
  and al.]{Lines2018}
Lines, S.; Mayne, N.J.; Boutle, I.A.; Manners, J.; Lee, G.K.H.; Helling, C.; Drummond, B.; Amundsen, D.S.; Goyal, J.; Acreman, D.M.; et al.
\newblock {Simulating the cloudy atmospheres of HD 209458 b and HD 189733 b
  with the 3D Met Office Unified Model}.
\newblock {\em A{\&}A} {\bf 2018}, {\em 615},~A97.

\bibitem[Miller-Ricci    {et~al.}(2009)Miller-Ricci, Seager, and
  Sasselov]{MillerRicci09}
Miller-Ricci, E.; Seager, S.; Sasselov, D.
\newblock {The Atmospheric Signatures of Super-Earths: How to Distinguish
  Between Hydrogen-Rich and Hydrogen-Poor Atmospheres}.
\newblock {\em Astrophys. J.} {\bf 2009}, {\em 690}.

\bibitem[Thorngren    {et~al.}(2019)Thorngren, Gao, and
  Fortney]{Thorngren2019a}
Thorngren, D.P.; Gao, P.; Fortney, J.J.
\newblock {The intrinsic temperature and radiative-convective boundary depth in
  the atmospheres of hot Jupiters}.
\newblock {\em Astrophys. J. Lett.} {\bf 2019}, {\em 884},~L6.

\bibitem[Podolak    {et~al.}(2019)Podolak, Helled, and Schubert]{Podolak2019}
Podolak, M.; Helled, R.; Schubert, G.
\newblock {Effect of non-adiabatic thermal profiles on the inferred composition
  of Uranus and Neptune}.
\newblock {\em Mon.  Not. R. Astron. Soc.} {\bf 2019}, {\em 487},~2653--2664.

\bibitem[Heng    {et~al.}(2012)Heng, Hayek, Pont, and Sing]{Heng2012}
Heng, K.; Hayek, W.; Pont, F.; Sing, D.K.
\newblock {On the effects of clouds and hazes in the atmospheres of hot
  Jupiters: Semi-analytical temperature-pressure profiles}.
\newblock {\em Mon.  Not. R. Astron. Soc.} {\bf 2012}, {\em 420},~20--36.

\bibitem[Linder    {et~al.}(2019)Linder, Mordasini, Molli{\`{e}}re, Marleau,
  Malik, Quanz, and Meyer]{Linder2018}
Linder, E.F.; Mordasini, C.; Molli{\`{e}}re, P.; Marleau, G.D.; Malik, M.;
  Quanz, S.P.; Meyer, M.R.
\newblock {Evolutionary models of cold and low-mass planets: Cooling curves,
  magnitudes, and detectability}.
\newblock {\em A\&A} {\bf 2019}, {\em 623},~A85.

\bibitem[{Kurosaki} and {Ikoma}(2017)]{Kurosaki2017}
{Kurosaki}, K.; {Ikoma}, M.
\newblock {Acceleration of Cooling of Ice Giants by Condensation in Early
  Atmospheres}.
\newblock {\em Astron. J.} {\bf 2017}, {\em 153},~260.

\bibitem[Barman    {et~al.}(2001)Barman, Hauschildt, Allard]{Barman2001}
{Barman}, T.S.; {Hauschildt}, P.H.; {Allard}, F.
\newblock {{Irradiated Planets}}.      %Added this and the following reference 
\newblock {\em Astrophys. J.} {\bf 2001}, {\em 556},~885--895.

\bibitem[Baraffe    {et~al.}(2003)Baraffe, Chabrier, Barman, Allard, Hauschildt]{Baraffe2003}
{Baraffe}, I.; {Chabrier}, G.; {Barman}, T.S.;  {Allard}, F.; {Hauschildt}, P.H.
\newblock {{Evolutionary models} for cool brown dwarfs and extrasolar giant planets. The case of HD 209458.}.
\newblock {\em A\&A} {\bf 2003}, {\em 402},~701--712.

\bibitem[Kataria    {et~al.}(2016)Kataria, Sing, Lewis, Visscher, Showman,
  Fortney, and Marley]{Kataria2016}
Kataria, T.; Sing, D.K.; Lewis, N.K.; Visscher, C.; Showman, A.P.; Fortney,
  J.J.; Marley, M.S.
\newblock {The atmospheric circulation of a nine-hot-Jupiter sample: Probing
  circulation and chemistry over a wide phase space}.
\newblock {\em Astrophys. J.} {\bf 2016}, {\em 821},~9.

\bibitem[{Molli{\`e}re}    {et~al.}(2017){Molli{\`e}re}, {van Boekel},
  {Bouwman}, {Henning}, {Lagage}, and {Min}]{Molliere2017}
{Molli{\`e}re}, P.; {van Boekel}, R.; {Bouwman}, J.; {Henning}, T.; {Lagage},
  P.O.; {Min}, M.
\newblock {Observing transiting planets with JWST. Prime targets and their
  synthetic spectral observations}.
\newblock {\em A\&A} {\bf 2017}, {\em 600},~A10.

\bibitem[Johnson    {et~al.}(2009)Johnson, Winn, Cabrera, and
  Carter]{Johnson2009}
Johnson, J.A.; Winn, J.N.; Cabrera, N.E.; Carter, J.A.
\newblock {A smaller radius for the transiting exoplanet WASP-10b}.
\newblock {\em Astrophys. J. Lett.} {\bf 2009}, {\em 692},~L100--L104.

\bibitem[Christian    {et~al.}(2009)Christian, Gibson, Simpson, Street,
  Skillen, Pollacco, {Collier Cameron}, Joshi, Keenan, Stempels, Haswell,
  Horne, Anderson, Bentley, Bouchy, Clarkson, Enoch, Hebb, H{\'{e}}brard,
  Hellier, Irwin, Kane, Lister, Loeillet, Maxted, Mayor, McDonald, Moutou,
  Norton, Parley, Pont, Queloz, Ryans, Smalley, Smith, Todd, Udry, West,
  Wheatley, and Wilson]{Christian2009}
Christian, D.J.; Gibson, N.P.; Simpson, E.K.; Street, R.A.; Skillen, I.;
  Pollacco, D.; {Collier Cameron}, A.; Joshi, Y.C.; Keenan, F.P.; Stempels,
  H.C.; et al.
\newblock {WASP-10b: A 3MJ, gas-giant planet transiting a late-type K star}.
\newblock {\em Mon.  Not. R. Astron. Soc.} {\bf 2009}, {\em 392},~1585.

\bibitem[Maciejewski    {et~al.}(2010)Maciejewski, Dimitrov, Neuh{\"{a}}user,
  Tetzlaff, Niedzielski, {St. Raetz}, Chen, Walter, Marka, Baar,
  Krejcov{\'{a}}, Budaj, Krushevska, Tachihara, Takahashi, and
  Mugrauer]{Maciejewski2011}
Maciejewski, G.; Dimitrov, D.; Neuh{\"{a}}user, R.; Tetzlaff, N.; Niedzielski,
  A.; {St. Raetz}; Chen, W.P.; Walter, F.; Marka, C.; Baar, S.;
et al.
\newblock {Transit timing variation and activity in the WASP-10 planetary
  system}.
\newblock {\em Mon.  Not. R. Astron. Soc.} {\bf 2010}, {\em 411}, 1204--1212.

\bibitem[Faedi    {et~al.}(2011)Faedi, Barros, Anderson, Brown, {Collier
  Cameron}, Pollacco, Boisse, H{\'{e}}brard, Lendl, Lister, Smalley, Street,
  Triaud, Bento, Bouchy, Butters, Enoch, Haswell, Hellier, Keenan, Miller,
  Moulds, Moutou, Norton, Queloz, Santerne, Simpson, Skillen, Smith, Udry,
  Watson, West, and Wheatley]{Faedi2011}
Faedi, F.; Barros, S.C.C.; Anderson, D.R.; Brown, D.J.A.; {Collier Cameron},
  A.; Pollacco, D.; Boisse, I.; H{\'{e}}brard, G.; Lendl, M.; Lister, T.A.;
 et al.
\newblock {WASP-39b: A highly inflated Saturn-mass planet orbiting a late
  G-type star}.
\newblock {\em A{\&}A} {\bf 2011}, {\em 531},~A40.

\bibitem[{Mordasini}    {et~al.}(2016){Mordasini}, {van Boekel},
  {Molli{\`e}re}, {Henning}, and {Benneke}]{Mordasini2016}
{Mordasini}, C.; {van Boekel}, R.; {Molli{\`e}re}, P.; {Henning}, T.;
  {Benneke}, B.
\newblock {The imprint of exoplanet formation history on observable present-day
  spectra of hot Jupiters}.
\newblock {\em Astrophys. J.} {\bf 2016}, {\em 832},~41.

\bibitem[{Lodders}(2003)]{Lodders2003}
{Lodders}, K.
\newblock {Solar system abundances and condensation temperatures of the
  elements}.
\newblock {\em Astrophys. J.} {\bf 2003}, {\em 591},~1220.

\bibitem[Saumon and Chabrier(1995)]{SaumonChabrier1995}
Saumon, D.; Chabrier, G.
\newblock {An equation of state for low-mass stars and giant planets}.
\newblock {\em  Astrophys.  J. Suppl.} {\bf 1995}, {\em 99},~713--741.

\bibitem[Hubbard and Marley(1989)]{Hubbard1989a}
Hubbard, W.B.; Marley, M.S.
\newblock {Optimized Jupiter, Saturn, and Uranus interior models}.
\newblock {\em Icarus} {\bf 1989}, {\em 78},~102.

\bibitem[Nettelmann    {et~al.}(2011)Nettelmann, Fortney, Kramm, and
  Redmer]{Nettelmann2011}
Nettelmann, N.; Fortney, J.J.; Kramm, U.; Redmer, R.
\newblock {Thermal evolution and structure models of the transiting super-Earth
  GJ 1214b}.
\newblock {\em Astrophys. J.} {\bf 2011}, {\em 733},~2.

\bibitem[Fortney and Nettelmann(2010)]{Fortney2010}
Fortney, J.J.; Nettelmann, N.
\newblock {The interior structure, composition, and evolution of giant
  planets}.
\newblock {\em Space Sci. Rev.} {\bf 2010}, {\em 152},~423--447.

\bibitem[Guillot(2010)]{Guillot2010}
Guillot, T.
\newblock {On the radiative equilibrium of irradiated planetary atmospheres}.
\newblock {\em A{\&}A} {\bf 2010}, {\em 520},~A27.

\bibitem[Marley    {et~al.}(1999)Marley, Gelino, Stephens, Lunine, and
  Freedman]{Marley1999}
Marley, M.S.; Gelino, C.; Stephens, D.; Lunine, J.I.; Freedman, R.
\newblock {Reflected spectra and albedos of extrasolar giant planets. I. Clear
  and cloudy atmospheres}.
\newblock {\em Astrophys. J.} {\bf 1999}, {\em 513},~879--893.

\bibitem[{Gelino}    {et~al.}(1999){Gelino}, {Marley}, {Stephens}, {Lunine},
  and {Freedman}]{Gelino1999}
{Gelino}, G.; {Marley}, M.; {Stephens}, D.; {Lunine}, J.; {Freedman}, R.
\newblock {Model Bond Albedos of Extrasolar Giant Planets}.
\newblock {\em Phys. Chem. Earth} {\bf 1999}, {\em 24},~573--578.

\bibitem[Sudarsky    {et~al.}(2000)Sudarsky, Burrows, and Pinto]{Sudarsky2000}
Sudarsky, D.; Burrows, A.; Pinto, P.
\newblock {Albedo and reflection spectra of extrasolar giant planets}.
\newblock {\em Am. Astron. Soc.} {\bf 2000}, {\em 538},~885--903.

\bibitem[Madhusudhan    {et~al.}(2014)Madhusudhan, Knutson, Fortney, and Barman]{Madhusudhan2014}
Madhusudhan, N.; Knutson, H.; Fortney, J.J.; Barman, T.
\newblock {{Exoplanetary atmospheres}}. In {\em Protostars and Planets VI 
  }; University of Arizona Press: Tucson, USA,  2014.

\bibitem[{Li}    {et~al.}(2018){Li}, {Jiang}, {West}, {Gierasch},
  {Perez-Hoyos}, {Sanchez-Lavega}, {Fletcher}, {Fortney}, {Knowles}, {Porco},
  {Baines}, {Fry}, {Mallama}, {Achterberg}, {Simon}, {Nixon}, {Orton},
  {Dyudina}, {Ewald}, and {Schmude}]{Li2018}
{Li}, L.; {Jiang}, X.; {West}, R.A.; {Gierasch}, P.J.; {Perez-Hoyos}, S.;
  {Sanchez-Lavega}, A.; {Fletcher}, L.N.; {Fortney}, J.J.; {Knowles}, B.;
  {Porco}, C.C.; {et al.}
\newblock {Less absorbed solar energy and more internal heat for Jupiter}.
\newblock {\em Nat. Commun.} {\bf 2018}, {\em 9},~3709.

\bibitem[Heng    {et~al.}(2014)Heng, Mendon{\c{c}}a, and Lee]{Heng2014}
Heng, K.; Mendon{\c{c}}a, J.M.; Lee, J.M.
\newblock {Analytical models of exoplanetary atmospheres. II. Radiative
  transfer via the two-stream approximation}.
\newblock {\em Astrophys.  J. Suppl.} {\bf 2014}, {\em 215},~4.

\bibitem[Fortney    {et~al.}(2007)Fortney, Marley, and Barnes]{Fortney2007}
Fortney, J.J.; Marley, M.S.; Barnes, J.W.
\newblock {Planetary radii across five orders of magnitude in mass and stellar
  insolation: Application to transits}.
\newblock {\em Astrophys. J.} {\bf 2007}, {\em 659},~1661--1672.

\bibitem[{Fortney}    {et~al.}(2006){Fortney}, {Saumon}, {Marley}, {Lodders},
  and {Freedman}]{Fortney2006}
{Fortney}, J.J.; {Saumon}, D.; {Marley}, M.; {Lodders}, K.; {Freedman}, R.
\newblock {Atmosphere, Interior, and Evolution of the Metal-rich Transiting
  Planet HD 149036b}.
\newblock {\em Astrophys. J.} {\bf 2006}, {\em 642},~495.

\bibitem[Molli{\`{e}}re    {et~al.}(2015)Molli{\`{e}}re, {Van Boekel},
  Dullemond, Henning, and Mordasini]{Molliere2015}
Molli{\`{e}}re, P.; {Van Boekel}, R.; Dullemond, C.; Henning, T.; Mordasini, C.
\newblock {Model atmospheres of irradiated exoplanets: The influence of stellar
  parameters, metallicity, and the C/O ratio}.
\newblock {\em Astrophys. J.} {\bf 2015}, {\em 813},~47.

\bibitem[Wakeford    {et~al.}(2017)Wakeford, Visscher, Lewis, Kataria, Marley,
  Fortney, and Mandell]{Wakeford2017}
Wakeford, H.R.; Visscher, C.; Lewis, N.K.; Kataria, T.; Marley, M.S.; Fortney,
  J.J.; Mandell, A.M.
\newblock {High-temperature condensate clouds in super-hot Jupiter
  atmospheres}.
\newblock {\em Mon. Not. R. Astron. Soc.} {\bf 2017}, {\em 464},~4247.

\bibitem[Valencia    {et~al.}(2013)Valencia, Guillot, Parmentier, and
  Freedman]{Valencia2013}
Valencia, D.; Guillot, T.; Parmentier, V.; Freedman, R.S.
\newblock {Bulk composition of GJ 1214b and other sub-neptune exoplanets}.
\newblock {\em Astrophys. J.} {\bf 2013}, {\em 775}.

\bibitem[Freedman    {et~al.}(2008)Freedman, Marley, and Lodders]{Freedman2008}
Freedman, R.; Marley, M.; Lodders, K.
\newblock {Line and mean opacities for ultracool dwarfs and extrasolar
  planets}.
\newblock {\em Astrophys.  J. Suppl.} {\bf 2008}, {\em 174},~504--513.

\bibitem[Marley(2000)]{Marley2000a}
Marley, M.S.
\newblock {The Role of Condensates in L- and T-dwarf Atmospheres}.
\newblock  In \emph{From Giant Planets to Cool Stars. ASP Conference Series}; ASP: San Francisco, CA, USA,  2000.%Newly added, please check if it is right. answer: correct

\bibitem[Lodders and Fegley(2006)]{Lodders2006}
Lodders, K.; Fegley, B.
\newblock {Chemistry of low mass substellar objects}. In {\em Astrophysics Update 
  2}; Springer: Heidelberg, DE,  2006.%Please add publisher’s location.answer: done

\bibitem[Morley    {et~al.}(2015)Morley, Fortney, Marley, Zahnle, Line,
  Kempton, Lewis, and Cahoy]{Morley2015}
Morley, C.V.; Fortney, J.J.; Marley, M.S.; Zahnle, K.; Line, M.; Kempton, E.;
  Lewis, N.; Cahoy, K.
\newblock {Thermal emission and reflected light spectra of super earths with
  flat transmission spectra}.
\newblock {\em Astrophys.  J.} {\bf 2015}, {\em 815},~110.

\bibitem[{Alexander} and {Armitage}(2009)]{Alexander2009}
{Alexander}, R.; {Armitage}, P.
\newblock {Giant planet migration, disk evolution, and the origin of
  transitional disks}.
\newblock {\em Astrophys.  J.} {\bf 2009}, {\em 704},~989.

\bibitem[Thorngren and Fortney(2018)]{Thorngren2018}
Thorngren, D.P.; Fortney, J.J.
\newblock {Bayesian Analysis of Hot-Jupiter Radius Anomalies: Evidence for
  Ohmic Dissipation?}
\newblock {\em Astron. J.} {\bf 2018}, {\em 155},~214.

\bibitem[Leconte    {et~al.}(2017)Leconte, Selsis, Hersant, and
  Guillot]{Leconte2017}
Leconte, J.; Selsis, F.; Hersant, F.; Guillot, T.
\newblock {Condensation-inhibited convection in hydrogen-rich atmospheres}.
\newblock {\em A \& A} {\bf 2017}, {\em 598},~A98.

\bibitem[Sing    {et~al.}(2016)Sing, Fortney, Nikolov, Wakeford, Kataria,
  Evans, Aigrain, Ballester, Burrows, Deming, D{\'{e}}sert, Gibson, Henry,
  Huitson, Knutson, Etangs, Pont, Showman, Vidal-Madjar, Williamson, and
  Wilson]{Sing2016}
Sing, D.K.; Fortney, J.J.; Nikolov, N.; Wakeford, H.R.; Kataria, T.; Evans,
  T.M.; Aigrain, S.; Ballester, G.E.; Burrows, A.S.; Deming, D.; et al.
\newblock {A continuum from clear to cloudy hot-Jupiter exoplanets without
  primordial water depletion}.
\newblock {\em Nature} {\bf 2016}, {\em 529},~59--62.

\bibitem[Nikolov    {et~al.}(2016)Nikolov, Sing, Gibson, Fortney, Evans,
  Barstow, Kataria, and Wilson]{Nikolov2016}
Nikolov, N.; Sing, D.K.; Gibson, N.P.; Fortney, J.J.; Evans, T.M.; Barstow,
  J.K.; Kataria, T.; Wilson, P.A.
\newblock {VLT FORS2 comparative transmission spectroscopy: Detection of Na in
  the atmosphere of WASP-39b from the ground}.
\newblock {\em Astrophys.  J.} {\bf 2016}, {\em 832},~191.

\bibitem[Becker    {et~al.}(2014)Becker, Lorenzen, Fortney, Nettelmann, Redmer,
  and Sch{\"{o}}ttler]{Becker2014}
Becker, A.; Lorenzen, W.; Fortney, J.J.; Nettelmann, N.; Redmer, R.;
  Sch{\"{o}}ttler, M.
\newblock {Ab initio equation of state for hydrogen (H-REOS.3) and helium
  (He-REOS.3) and their implications for the interior of brown dwarfs}.
\newblock {\em Astrophys.  J. Suppl.} {\bf 2014}, {\em 215},~A21.

\bibitem[Chabrier and Baraffe(2007)]{Chabrier2007}
Chabrier, G.; Baraffe, I.
\newblock {Heat transport in giant (exo)planets: A new perspective}.
\newblock {\em  Astrophys.  J. Lett.} {\bf 2007}, {\em 661},~L81--L84.

\bibitem[Leconte and Chabrier(2013)]{Leconte2013}
Leconte, J.; Chabrier, G.
\newblock {Layered convection as the origin of Saturn's luminosity anomaly}.
\newblock {\em Nat. Geosci.} {\bf 2013}, {\em 6},~347--350.

\bibitem[{Malik}    {et~al.}(2019){Malik}, {Kitzmann}, {Mendon\c{c}a}, {Grimm},
  {Marleau}, {Linder}, {Tsai}, and {Heng}]{Malik2019}
{Malik}, M.; {Kitzmann}, D.; {Mendon\c{c}a}, J.; {Grimm}, S.; {Marleau}, G.D.;
  {Linder}, E.; {Tsai}, S.M.; {Heng}, K.
\newblock {Self-luminous and Irradiated Exoplanetary Atmospheres Explored with
  HELIOS}.
\newblock {\em Astron. J.} {\bf 2019}, {\em 157},~170.

\bibitem[{Allard}    {et~al.}(2001){Allard}, {Hauschildt}, {Alexander},
  {Tamanai}, and {Schweitzer}]{Allard2001}
{Allard}, F.; {Hauschildt}, P.; {Alexander}, D.; {Tamanai}, A.; {Schweitzer},
  A.
\newblock {The Limiting Effects of Dust in Brown Dwarf Model Atmospheres}.
\newblock {\em Astrophys.  J.} {\bf 2001}, {\em 556},~357.

\bibitem[{Hauschildt} and {Baron}(1999)]{Hauschildt1999}
{Hauschildt}, P.; {Baron}, E.
\newblock {Numerical Solution of the Expanding Stellar Atmosphere Problem}.
\newblock {\em J. Comput. Appl. Math.} {\bf 1999}, {\em 109},~41.

\bibitem[{Ackerman} and {Marley}(2001)]{Ackerman2001}
{Ackerman}, A.; {Marley}, M.
\newblock {Precipitating condensation clouds in substellar atmospheres}.
\newblock {\em Astrophys.  J.} {\bf 2001}, {\em 556},~872.

\end{thebibliography}
\end{document}